\documentclass[12pt]{article}
\usepackage{latexsym,amssymb,amsfonts,amsmath}
\usepackage{mathrsfs}
\usepackage[dvips]{graphicx}
\newlength{\extraspace}
\newlength{\extraspaces}
\newcommand{\be}{\begin{equation}
\addtolength{\abovedisplayskip}{\extraspaces}
\addtolength{\belowdisplayskip}{\extraspaces}
\addtolength{\abovedisplayshortskip}{\extraspace}
\addtolength{\belowdisplayshortskip}{\extraspace}}
\newcommand{\ee}{\end{equation}}

\newcommand{\bea}{\begin{eqnarray}
\addtolength{\abovedisplayskip}{\extraspaces}
\addtolength{\belowdisplayskip}{\extraspaces}
\addtolength{\abovedisplayshortskip}{\extraspace}
\addtolength{\belowdisplayshortskip}{\extraspace}}
\newcommand{\eea}{\end{eqnarray}}

\DeclareMathOperator{\Tr}{Tr} 
\DeclareMathOperator{\diag}{diag} 
\DeclareMathOperator{\re}{Re}

\newcommand{\Overline}[2][1]{%
  {}\mkern#1mu \overline{\mkern-#1mu #2 \mkern-#1mu}\mkern#1mu {}}

\begin{document}

\addtolength{\baselineskip}{.8mm}

{\thispagestyle{empty}

\noindent \hspace{1cm} \hfill IFUP--TH/2018 \hspace{1cm}\\

\begin{center}
{\Large\bf Study of the theta dependence of the vacuum energy}\\
{\Large\bf density in chiral effective Lagrangian models}\\
{\Large\bf at zero temperature}\\
\vspace*{1.0cm}
{\large
Francesco Luciano\footnote{E-mail: f.luciano@live.it}
and
Enrico Meggiolaro\footnote{E-mail: enrico.meggiolaro@unipi.it}
}\\
\vspace*{0.5cm}{\normalsize
{Dipartimento di Fisica, Universit\`a di Pisa,
and INFN, Sezione di Pisa,\\
Largo Pontecorvo 3, I-56127 Pisa, Italy}}\\
\vspace*{2cm}{\large \bf Abstract}
\end{center}

\noindent
We perform a systematic study of the modifications to the
QCD vacuum energy density $\epsilon_{vac}$ in the zero-temperature case ($T=0$)
caused by a small, but non-zero, value of the parameter $\theta$,
using different effective Lagrangian models which include the flavour-singlet meson field and implement the $U(1)$ axial anomaly of the fundamental theory.
In particular, we derive the expressions for the topological susceptibility $\chi$ and for the second cumulant $c_4$ starting from the $\theta$ dependence of $\epsilon_{vac}(\theta)$ in the various models that we have considered.
Moreover, we evaluate numerically our results, so as to compare them with each other, with the predictions of the Chiral Effective Lagrangian, and, finally, also with the available lattice data.

}
\newpage

\section{Introduction}

The discovery of instantons in the $70$'s \cite{BPST1975} made clear that topology was a relevant aspect of the dynamics of the low-energy degrees of freedom in QCD \cite{tHooft1976,Witten1979,Veneziano1979},
but it also raised another important issue: if one introduces in the QCD Lagrangian an additional term $\mathscr{L}_\theta = \theta Q$,
where $Q(x)=\frac{g^{2}}{64\pi^{2}}\varepsilon^{\mu\nu\rho\sigma} F_{\mu\nu}^{a}(x)F_{\rho\sigma}^{a}(x)$ is the so-called \emph{topological charge density},
despite the fact that $Q = \partial^\mu K_\mu$, where $K^\mu$ is the so-called
\emph{Chern-Simons current}, its contribution in the quantum theory would be non-zero thanks to the existence of configurations with non-trivial topology (such as instantons). This term, usually referred to as \emph{topological term} or as \emph{$\theta$-term} (from the name of the coefficient that appears in front of it), is particularly interesting since it introduces an explicit breaking of the CP symmetry in QCD (referred to as \emph{strong-CP violation}), absent in the original theory.
So far, however, no violation of the CP symmetry in strong interactions has been observed experimentally, so that the parameter $\theta$ is believed to be zero (or ``practically'' zero), despite the fact that it could assume, in principle, whatever value in $[0,2\pi)$.
In particular, one can find a relation between the magnitude of the parameter $\theta$ and the neutron electric-dipole moment \cite{Weinberg-book},
$d_N \simeq \frac{M_\pi^2}{M_N^3}\, e \left|\theta \right| \simeq 10^{-16} \left|\theta \right| \; e\cdot cm$,
where $M_N$ is the neutron mass, while $M_\pi$ is the pion mass. From the experimental data \cite{Neutron-EDM} we know that $d_N<10^{-26} e\cdot cm$, which leads to an upper bound:
\begin{equation}\label{upper bound on theta}
\left|\theta \right| < 10^{-10} .
\end{equation}
(More refined relations among the neutron electric dipole moment and the $\theta$ angle were derived by Baluni \cite{Baluni1979}, in the framework of the so-called \emph{bag model}, by Crewther, Di Vecchia, Veneziano, and Witten \cite{CDVW1979}, using the Chiral Perturbation Theory, and by many others using different approaches: see Sec. 7.1 of Ref. \cite{VP2009} for a more detailed discussion
and also Ref. \cite{EDM-lattice} for a recent lattice determination.)

This ``fine-tuning'' problem (usually referred to as the \emph{strong-CP problem}), is still an open issue, despite possible solutions have been proposed
(the most famous one being that of Peccei and Quinn \cite{PQ1977}, who proposed a mechanism, based on a new $U(1)$ symmetry and involving a new light pseudoscalar particle called \emph{axion} \cite{WW1978}, in order to dynamically rotate away the $\theta$-dependence of the theory).

However, it is nonetheless interesting to study the dependence of QCD on finite $\theta$: the insertion of the topological term with $\theta\neq 0$ in the QCD Lagrangian causes (by virtue of the non-trivial topology) a modification of the partition function of the theory and, therefore, a non-trivial dependence on $\theta$ of the \emph{vacuum energy density} $\epsilon_{vac}(\theta)$, which will be the object of our investigations in this paper.

Let us write explicitly the expression for the partition function of the theory with $N_f$ quark flavours and with the inclusion of the $\theta$-term:
\begin{equation}\label{total partition function}
Z = \int [dA][d\Overline[2]{q} dq]e^{i\int d^4x \mathscr{L}_{tot}} ,
\end{equation}
where $\mathscr{L}_{tot}= -\frac{1}{2}\Tr\left[F_{\mu\nu}F^{\mu\nu}\right] +i\Overline[2]{q}\gamma^\mu D_\mu q-\Overline[2]{q}_R \mathcal{M} q_L -\Overline[2]{q}_L \mathcal{M}^{\dagger} q_R +\theta Q$, with $\mathcal{M}$ a general complex mass matrix for the quarks.
If we now perform a change of the (dummy) fermionic integration variables in
\eqref{total partition function} in the form of a $SU(N_f)_L \otimes SU(N_f)_R \otimes U(1)_A$ transformation,\footnote{Throughout this paper, we shall use the following notations for the left-handed and right-handed quark fields: $q_{L,R} \equiv \frac{1}{2} (1 \pm \gamma_5) q$, with $\gamma_5 \equiv -i\gamma^0\gamma^1\gamma^2\gamma^3$. Moreover, we shall adopt the convention $\varepsilon^{0123} = -\varepsilon_{0123} = +1$ for the (Minkowskian) completely antisymmetric tensor $\varepsilon^{\mu\nu\rho\sigma} (=-\varepsilon_{\mu\nu\rho\sigma})$ which appears in the expression of the topological charge density $Q(x)$.}
\begin{equation} \label{transformation SU(L)_L x SU(L)_R x U(1)_A}
\left\{
\begin{aligned}
q_L & \rightarrow q_L' = \tilde{V_L} q_L = e^{i\alpha} V_L q_L ,\\ 
q_R & \rightarrow q_R' = \tilde{V_R} q_R = e^{-i\alpha} V_R q_R ,
\end{aligned}
\right.
\end{equation}
where $V_L,~V_R \in SU(N_f)$, we see that, because of the non-invariance of the fermionic functional-integral measure
($[d\Overline[2]{q} dq]\rightarrow [d\Overline[2]{q}' dq']=[d\Overline[2]{q} dq] e^{-i2L\alpha \int d^4x \, Q(x)}$)
and of the mass term,
the partition function $Z$ is invariant under the following changes:
\begin{equation}\label{changes in Z}
\left\{
\begin{aligned}
\mathcal{M}&\rightarrow \mathcal{M}'=\tilde{V}_R^{\dagger}\, \mathcal{M} \, \tilde{V}_L , \\
\theta & \rightarrow \theta' = \theta - 2L \alpha .
\end{aligned}
\right.
\end{equation}
We immediately notice that, if $\mathcal{M}$ is invertible ($\det\mathcal{M}\neq 0$), we have:
$\arg(\det\mathcal{M}')=\arg(\det\mathcal{M})+2L \alpha$,
so that, under the transformation \eqref{transformation SU(L)_L x SU(L)_R x U(1)_A}-\eqref{changes in Z}, the following combination:
\begin{equation}\label{physical theta}
\theta_{phys} \equiv \theta + \arg(\det\mathcal{M})
\end{equation}
stays unchanged. This is the ``physical'' value of the parameter $\theta$:
a non-zero value of $\theta_{phys}$ implies a strong CP-violation and
the upper bound \eqref{upper bound on theta} actually refers to $\theta_{phys}$.

Eqs. \eqref{changes in Z} and \eqref{physical theta} also imply that, if the mass matrix is invertible, then it is possible to move all the dependence on the parameter $\theta$ into the mass term. In fact, performing a transformation \eqref{transformation SU(L)_L x SU(L)_R x U(1)_A}-\eqref{changes in Z} with
$\alpha=\frac{\theta}{2L}$, we obtain $\theta'=0$ and $\arg(\det\mathcal{M}') = \theta_{phys}$.
On the other hand, if we take $\mathcal{M}$ to coincide with the \emph{physical} quark-mass matrix $M \equiv \diag(m_1,\ldots m_{N_f})$, with $m_i \in \mathbb{R}^+ \; \forall i$ (which is always possible, by means of a transformation
\eqref{transformation SU(L)_L x SU(L)_R x U(1)_A}-\eqref{changes in Z}),
we have $\arg(\det M)=0$ and $\theta = \theta_{phys}$. (Of course, if at least one quark is massless, we have $\det {M}=0$ and, in this case, it is possible to rotate away all the dependence on $\theta_{phys}$ from the theory.)

From now on, we shall consider the partition function $Z[\theta]$ in this case
($\mathcal{M} = M$ and $\theta = \theta_{phys}$).
In particular, we are interested in the $\theta$-dependence of the vacuum energy density $\epsilon_{vac}(\theta)$, which is related to the partition function $Z[\theta]$ by the following well-known relation:
\begin{equation}\label{vacuum energy density definition}
Z[\theta]\equiv \frac{1}{\mathcal{N}}e^{-i\Omega \epsilon_{vac}(\theta)} \Rightarrow \epsilon_{vac}(\theta) = \frac{i}{\Omega}\log Z[\theta] + const.,
\end{equation}
where $\mathcal{N}$ is a normalizing constant while $\Omega=VT$ is the four-volume considered (sending $\Omega \to \infty^4$ at the end).\footnote{The expression \eqref{vacuum energy density definition} is referred to the partition function of the theory in the Minkowski space-time. It is also common to express it in terms of the partition function $Z_E[\theta]$ of the theory in the Euclidean space-time as follows: $\epsilon_{vac}(\theta) = -\frac{1}{\Omega_E}\log Z_E[\theta] + const.$, where $\Omega_E=VT_E$ is the Euclidean four-volume, with Euclidean time $T_E$}
Being $\theta$ very small, it makes sense to Taylor-expand the vacuum energy density around $\theta=0$:
\begin{equation}\label{Taylor expansion of vacuum energy density}
\epsilon_{vac}(\theta) = \epsilon_{vac}(0) +  \frac{1}{2!} c_2 \theta^2 + \frac{1}{4!} c_4 \theta^4 + \ldots ~,~~~ \text{with:}~~
c_n \equiv \left.\frac{\partial^n \epsilon_{vac}(\theta)}{\partial \theta^n}\right|_{\theta=0} .
\end{equation}
Only even powers of $\theta$ appear in \eqref{Taylor expansion of vacuum energy density} since the coefficients $c_n$ of the odd-power terms vanish by parity-invariance at $\theta=0$. The coefficients $c_n$ of this expansion are related to the correlation functions of the topological charge density at $\theta=0$.
More explicitly, starting from the expression \eqref{vacuum energy density definition} and indicating with $Q_{tot} \equiv \int d^4 x\, Q(x)$ the (total) topological charge, one easily finds that:
\begin{equation}\label{topological susceptibility from vacuum energy density}
c_2 \equiv \left.\frac{\partial^2 \epsilon_{vac}(\theta)}{\partial \theta^2}\right|_{\theta=0} = -\frac{i}{\Omega} \langle Q_{tot}^2 \rangle_{\theta=0} \equiv \chi,
\end{equation}
i.e., the coefficient $c_2$ of the $\theta^2$ term in \eqref{Taylor expansion of vacuum energy density} coincides with the so-called \emph{topological susceptibility} of the theory at $\theta=0$:
$\chi \equiv -\frac{i}{\Omega} \langle Q_{tot}^2 \rangle_{\theta=0} =
-i \int d^4 x \langle T Q(x) Q(0) \rangle_{\theta=0}$.

Concerning the coefficient $c_4$, it turns out to coincide with the \emph{second cumulant} of the probability distribution of the topological charge-density operator $Q$ \cite{VP2009}:
\begin{equation}\label{c4 from vacuum energy density}
c_4 \equiv \left.\frac{\partial^4 \epsilon_{vac}(\theta)}{\partial \theta^4}\right|_{\theta=0} = \frac{i}{\Omega}\left(\langle Q_{tot}^4 \rangle_{\theta=0} - 3 \langle Q_{tot}^2 \rangle^2_{\theta=0}\right) ,
\end{equation}
which is related to the $\eta'-\eta'$ elastic scattering amplitude \cite{Veneziano1979} and to the non-gaussianity of the topological charge distribution \cite{VP2009}.

Therefore, the expansion \eqref{Taylor expansion of vacuum energy density} can be rewritten as:
\begin{equation}\label{susceptibility and c4 in vacuum energy density}
\epsilon_{vac}(\theta)=\epsilon_{vac}(0)+\frac{1}{2}\chi \theta^2 + \frac{1}{24}c_4 \theta^4+\ldots
\end{equation}
The strategy of this paper consists in computing the dependence on $\theta$ of the vacuum energy density, so as to obtain, exploiting the relations \eqref{topological susceptibility from vacuum energy density} and \eqref{c4 from vacuum energy density}, the expressions of the topological susceptibility $\chi$ and of the second cumulant $c_4$ in terms of the fundamental parameters of the theory,
not using directly the \emph{fundamental} theory (which is anyhow possible using its formulation on the lattice: see the discussion in Sec. 6),
but using some relevant \emph{effective} Lagrangian models.

We shall first consider, in Sec. 2, the \emph{Chiral Effective Lagrangian} in the case of $L$ ($\le N_f$) \emph{light} quark flavours (taken to be massless in the ideal \emph{chiral} limit): the physically relevant cases are $L=2$, with the
quarks $up$ and $down$, and $L=3$, including also the $strange$ quark \cite{Weinberg1967,Weinberg1979,GL1982-1984,GL1985}.
This effective theory describes the low-energy dynamics for the lightest hadronic states in the spectrum of QCD, i.e., the lightest non-flavour-singlet pseudoscalar mesons, which are identified with the $L^2-1$ pseudo-Goldstone bosons originated by the spontaneous breaking of the $SU(L)_L \otimes SU(L)_R$ chiral symmetry.
The results that we shall report in Sec. 2 are already well known in the literature (see, in particular, Refs. \cite{Smilga-book,MC2009,GM2015,GHVV2016}).
However, for the benefit of the reader, we have decided to report here some details of the calculations of $\chi$ and $c_4$ also in this case since this will allow us to introduce the basic notations and the main techniques for performing the calculations in the other cases.
Moreover, this case is an important frame of reference for the other effective models that we shall discuss in the rest of the paper.

In Secs. 3 and 4 we shall consider different effective Lagrangian models which include the flavour-singlet meson field and also implement the $U(1)$ axial anomaly of the fundamental theory.
In the last decades there were essentially two different ``schools of thought'' debating on how to address this issue: the first assumes that the dominant fluctuations are semiclassical instantons, while the second is based upon the large-$N_c$ limit of a $SU(N_c)$ gauge theory, and assumes that the dominant fluctuations are not semiclassical but quantum. The two models that we shall consider in Secs. 3 and 4 belong respectively to the first trend (the so-called \emph{Extended (Non-)Linear sigma model} \cite{ELSM1,ELSM2,ELSM3}) and to the second one (the model of Witten, Di Vecchia, Veneziano, \emph{et al.} \cite{WDV1,WDV2,WDV3}).

In Sec. 5, we shall consider another effective Lagrangian model (which
was originally proposed in Refs. \cite{EM1994} and elaborated on in Refs.
\cite{MM2003,EM2011,MM2013}), which is in a sense in-between the \emph{Extended (Non-)Linear sigma model} and the model of Witten, Di Vecchia, Veneziano, \emph{et al.}: for this reason we shall call it the \emph{Interpolating model}.

Finally, in Sec. 6 we shall draw our conclusions, summarizing the analytical results that we have obtained for the topological susceptibility $\chi$ and the second cumulant $c_4$ in the four different frameworks mentioned above and also evaluating numerically our results, so as to critically compare them with each other and with the available lattice results.

\section{The Chiral Effective Lagrangian}

We first consider the Chiral Effective Lagrangian in the case of $L$ light quark flavours: the results that we shall report in this section are already well known in the literature (see, in particular, Refs. \cite{Smilga-book,MC2009,GM2015,GHVV2016}).
However, for the benefit of the reader, we have decided to report here some details of the calculations of $\chi$ and $c_4$ also in this case since this will allow us to introduce the basic notations and the main techniques for performing the calculations in the other cases.
Moreover, this case is an important frame of reference for all the other models that we shall discuss: in fact, if one ``neglects'' the presence of the flavour-singlet meson field and of the $U(1)$ axial anomaly (formally sending its mass to infinity), all the predictions derived in the other models must reduce to those that will be found in this section.

The chiral effective Lagrangian formulation was introduced by Weinberg \cite{Weinberg1967}
and was later elaborated on, becoming one of the most important tool to investigate the dynamics of the effective degrees of freedom of the low-energy regime of QCD \cite{Weinberg1979,GL1982-1984,GL1985}.
The idea carried on by Weinberg \emph{et al.} was that of building an effective theory for the lightest hadronic states in the spectrum of the theory, i.e., the lightest pseudoscalar mesons, which are the pseudo-Goldstone bosons originated by the spontaneous breaking of the chiral symmetry.
This purpose can be achieved by writing down all the terms consistent with the symmetries of the fundamental theory, thereby obtaining an ``exact'' theory. However, the number of terms which satisfy the requirement is infinite: so, in order to be able to make any definite physical prediction, it is necessary to endow the theory with a power-counting ordering scheme which organizes the terms, providing a criterion to decide whether to keep or not a term at a given order.
Such a criterion is the low-energy expansion, or the \emph{$p$-expansion}: it consists in sorting the terms of the Chiral Effective Lagrangian on the basis of their number of derivatives, i.e., for the amplitudes in momentum space, on their order in the momentum-scale $p$.
So, a generic Chiral Effective Lagrangian is written as:
\begin{equation}\label{Generic Chiral Effective Lagrangian}
\mathscr{L}_{eff}= \mathscr{L}_{eff}^{(0)} + \mathscr{L}_{eff}^{(2)} + \mathscr{L}_{eff}^{(4)} + \mathscr{L}_{eff}^{(6)} + \ldots ,
\end{equation}
where $\mathscr{L}_{eff}^{(2n)}$ gathers all the terms of order $p^{2n}$ (i.e, with $2n$ derivatives, the quark-mass matrix $\mathcal{M}$ counting as $p^2$, i.e., as two derivatives), while the odd-power terms are ruled out by Lorentz invariance. The term $\mathscr{L}_{eff}^{(0)}$ turns out to be an irrelevant constant, which can be neglected.
In this paper, we shall make use of the Chiral Effective Lagrangian at the lowest (leading) nontrivial order $\mathcal{O}(p^2)$. Here, we limit ourselves to report the final result (for a dissertation on the Chiral Effective Lagrangian up to the next-to-leading order $\mathcal{O}(p^4)$, see Ref. \cite{GL1985}):
\begin{equation}\label{Chiral Effective Lagrangian O(p^2)}
\mathscr{L}_{eff}^{(2)}(U,U^{\dagger})= \frac{1}{2}\Tr [\partial_\mu U \partial^\mu U^{\dagger}] + \frac{B_m}{2\sqrt{2}}\Tr\left[\mathcal{M} U +\mathcal{M}^\dagger U^{\dagger}\right] ,
\end{equation}
where:
\begin{itemize}
\item the field $U$, describing only the $L^2-1$ non-flavour-singlet pseudo-Goldstone bosons, is an element of the group $SU(L)$, up to a multiplicative constant. In other words, it can be written as:
\begin{equation}
U\equiv \frac{F_\pi}{\sqrt{2}}\, U'~,~~~ U'\in SU(L) ,
\end{equation}
where $F_\pi$ is the usual \emph{pion decay constant};
\item $\mathcal{M}$ is a complex quark-mass matrix, which, considering the relation \eqref{physical theta} between the coefficient $\theta$ of the topological term and the argument of the determinant of the mass matrix, can be taken to be:
\begin{equation}\label{mass matrix with theta term}
\mathcal{M}=Me^{i\frac{\theta_{phys}}{L}} ,
\end{equation}
where $M=\diag (m_1,\ldots , m_L)$ is the physical (real and diagonal) quark-mass matrix. In this way, we are moving all the dependence on $\theta_{phys}$ into the mass term. In order to simplify the notation, from now on we shall write $\theta$ in place of $\theta_{phys}$;
\item $B_m$ is a constant having the dimension of an energy squared, often written as:
\begin{equation}\label{B_m definition}
B_m = 2 F_\pi B ,
\end{equation}
where $B$ is a constant, carrying the dimension of an energy, which relates the mass of the quarks \emph{up} and \emph{down} to the mass of the pions through: $M_\pi^2 = B(m_u + m_d)$.
\end{itemize}
We can rewrite the Chiral Effective Lagrangian
\eqref{Chiral Effective Lagrangian O(p^2)} as:
\begin{equation}\label{Chiral Effective Lagrangian O(p^2) bis}
\mathscr{L}_{eff}^{(2)}(U,U^{\dagger})= \frac{1}{2}\Tr [\partial_\mu U \partial^\mu U^{\dagger}] - V(U,U^\dagger) ,
\end{equation}
where the potential $V$ is given by:
\begin{equation}\label{Eff ch Lagr: potential}
V(U,U^\dagger) = -\frac{B_m}{2\sqrt{2}}\Tr\left[\mathcal{M}U +\mathcal{M}^{\dagger} U^{\dagger}\right] = -\frac{B_m}{\sqrt{2}} \re\left[\Tr\left(Me^{i\theta /L}U\right)\right] .
\end{equation}
We shall use the fact that (up to an irrelevant constant with respect to $\theta$), the vacuum energy density $\epsilon_{vac}(\theta)$ coincides with the \emph{minimum} of the potential $V$ obtained with a configuration of fields constant with respect to space-time coordinates $x$ (see Refs. \cite{Smilga-book,LM1992} and references therein):
\begin{equation}\label{vacuum energy density = V_min}
\epsilon_{vac}(\theta) \simeq V_{min}(\theta) + const.
\end{equation}
Given that we are considering $M=\diag(m_1,\ldots , m_L)$, it is reasonable to look for the minimum of the potential guessing a configuration of the field $U$ in a diagonal form.
So, being, in this case, $U=\frac{F_\pi}{\sqrt{2}}\, U'$, where $U'$ is an element of $SU(L)$, we set:
\begin{equation}\label{Diagonal form of U}
U = \frac{F_\pi}{\sqrt{2}}\diag \left( e^{i\alpha_1},\ldots , e^{i\alpha_L}\right) ,
\end{equation}
where the $\alpha_j$ are constant phases, satisfying the constraint:
\begin{equation}\label{Constraint of the determinant of U}
\det U' = e^{i\sum_j \alpha_j} = 1 \Longrightarrow \sum\limits_{j=1}^L \alpha_j = 0 .
\end{equation}
Substituting the explicit expressions for $M$ and $U$ into Eq. \eqref{Eff ch Lagr: potential}, we find:
\begin{equation}\label{Eff ch Lagr: explicit potential}
V = -\frac{F_\pi B_m}{2} \sum\limits_{j=1}^L m_j \cos\phi_j ,
\end{equation}
where we have defined $\phi_j \equiv \frac{\theta}{L}+\alpha_j$.
Starting from Eq. \eqref{Constraint of the determinant of U}, we see that the phases $\phi_j$ must satisfy the constraint:
\begin{equation}\label{Constraint on the phi}
\sum\limits_{j=1}^L \phi_j = \sum\limits_{j=1}^L \left(\frac{\theta}{L}+\alpha_j\right) = \theta .
\end{equation}
It is now more convenient to consider separately the special case $L=2$ and the more general case $L \ge 2$: in fact, the former can be easily solved exactly, for any values of $\theta$ and of the quark masses; on the contrary, the latter cannot be solved exactly (in ``closed form'') in general, but only an approximate solution can be derived.

\subsection{A special case: $L=2$}
In this case, it is easy to find the explicit expressions of the phases $\phi_1$ and $\phi_2$ which minimize the potential \eqref{Eff ch Lagr: explicit potential}, with the constraint \eqref{Constraint on the phi}:
\begin{equation}\label{Solution of the minimization in simple model L=2}
\phi_1 = \arctan\left(\frac{m_2\sin\theta}{m_1+m_2\cos\theta}\right) ~,~~~
\phi_2=\theta - \phi_1 .
\end{equation}
Substituting \eqref{Solution of the minimization in simple model L=2} in \eqref{Eff ch Lagr: explicit potential}, the following expression for the minimum of the potential is found:
\begin{equation}\label{Eff ch Lagr: final form of the potential for L=2}
V(\theta)=\epsilon_{vac}(\theta)=-\frac{F_\pi B_m}{2}\sqrt{m_1^2 + m_2^2 + 2m_1 m_2 \cos\theta} .
\end{equation}
In the end, we are able to find the expressions for the topological susceptibility $\chi$ and the second cumulant $c_4$ \cite{Smilga-book,MC2009,GM2015,GHVV2016}:
\begin{equation}\label{Eff ch Lagr: topological susceptibility for L=2}
\chi=\left.\frac{\partial^2 \epsilon_{vac}(\theta)}{\partial\theta^2}\right|_{\theta=0} = \frac{F_\pi B_m}{2}\left(\frac{1}{m_1}+\frac{1}{m_2}\right)^{-1} ,
\end{equation}
\begin{equation}\label{Eff ch Lagr: second cumulant for L=2}
c_4 = \left.\frac{\partial^4 \epsilon_{vac}(\theta)}{\partial\theta^4}\right|_{\theta=0} = -\frac{F_\pi B_m}{2}\left(\frac{1}{m_1^3} + \frac{1}{m_2^3}\right)\left(\frac{1}{m_1} + \frac{1}{m_2}\right)^{-4} .
\end{equation}

\subsection{The more general case: $L\ge 2$}

In the more general case $L\ge 2$ it is not possible to find an exact analytical solution, as in the previous case. However, given that our final purpose is to obtain the expressions for $\chi$ and $c_4$, which are by definition evaluated at $\theta=0$, we can implement a Taylor expansion of the potential around $\theta=0$.
If we set $\theta=0$, it is easy to show that the form of the field $U$ which minimizes the potential is $U=\frac{F_\pi}{\sqrt{2}}\mathbf{I}$.
We can thus implement a Taylor expansion of the potential \eqref{Eff ch Lagr: explicit potential} considering both $\theta\ll 1$ and $\phi_i \ll 1$ $\forall i$.
After some calculations, the following expression for the phases $\phi_i$ which minimize the potential \eqref{Eff ch Lagr: explicit potential}, with the constraint \eqref{Constraint on the phi}, is found:
\begin{equation}\label{Eff ch Lagr: final form of phi}
\phi_i=\frac{\bar{m}}{m_i}\theta + \frac{1}{6} \frac{\bar{m}}{m_i} \left[\left(\frac{\bar{m}}{m_i}\right)^2 - \sum\limits_{j=1}^L\left(\frac{\bar{m}}{m_j}\right)^3\right]\theta^3+\mathcal{O}(\theta^5) ,
\end{equation}
where we have defined:
\begin{equation}\label{m-bar}
\bar{m}\equiv\left(\sum\limits_{i=1}^L \frac{1}{m_i} \right)^{-1} .
\end{equation}
Finally, inserting \eqref{Eff ch Lagr: final form of phi} in
\eqref{Eff ch Lagr: explicit potential}, we find:
\begin{equation}\label{Eff ch Lagr: final form of the potential L>2}
V(\theta)=\epsilon_{vac}(\theta)= const. + \frac{1}{2} \left[ \frac{F_\pi B_m\bar{m}}{2} \right] \theta^2 + \frac{1}{24} \left[ -\frac{F_\pi B_m\bar{m}}{2}\sum_{j=1}^L\left(\frac{\bar{m}}{m_j}\right)^3 \right] \theta^4 + \ldots
\end{equation}
From this expression, we extract the final results for the topological susceptibility and for the second cumulant \cite{Smilga-book,MC2009,GM2015,GHVV2016}:
\begin{equation}\label{Eff ch Lagr: topological susceptibility for L>2}
\chi=\frac{F_\pi B_m\bar{m}}{2}=\frac{F_\pi B_m}{2}\left(\sum\limits_{j=1}^L \frac{1}{m_j}\right)^{-1} ,
\end{equation}
\begin{equation}\label{Eff ch Lagr: second cumulant for L>2}
c_4=-\frac{F_\pi B_m\bar{m}}{2}\sum\limits_{j=1}^L\left(\frac{\bar{m}}{m_j}\right)^3 = -\frac{F_\pi B_m}{2}\left(\sum\limits_{j=1}^L \frac{1}{m_j}\right)^{-4}\sum\limits_{j=1}^L\frac{1}{m_j^3} .
\end{equation}
These expressions correctly reduce to \eqref{Eff ch Lagr: topological susceptibility for L=2}-\eqref{Eff ch Lagr: second cumulant for L=2} if the number of light flavours considered is set to $L=2$.
In this respect, we want also to oberve that, if one of the quark masses, let's say $m_L$, is much larger than the other masses $m_1,\ldots,m_{L-1}$, we can formally take the limit $m_L \to \infty$ in the expressions
\eqref{Eff ch Lagr: topological susceptibility for L>2} and
\eqref{Eff ch Lagr: second cumulant for L>2} for $\chi^{(L)}$ and $c_4^{(L)}$,
which then reduce to $\chi^{(L-1)}$ and $c_4^{(L-1)}$, respectively.
In the real-world case, for example, the mass of the \emph{strange} quark,
$m_s$, is much larger than the masses $m_u$ and $m_d$ of the \emph{up} and
\emph{down} quarks: for this reason, in Sec. 6 we shall evaluate numerically
the expressions \eqref{Eff ch Lagr: topological susceptibility for L>2} and
\eqref{Eff ch Lagr: second cumulant for L>2} both for the case $L=2$, with
only the quarks \emph{up} and \emph{down}, and for the case $L=3$, where also
the \emph{strange} quark is taken into account. 

\subsection{Considerations on the results}

We recall that, if at least one quark is massless, the partition function of the theory (and, so, the vacuum energy density) turns out to be independent of $\theta$: we thus expect that, being the topological susceptibility and the second cumulant derivatives of the vacuum energy density with respect to $\theta$, if we let one of the quark masses tend to zero, both $\chi$ and $c_4$ will tend to zero as well. It is easy to check that the expressions
\eqref{Eff ch Lagr: topological susceptibility for L>2} and
\eqref{Eff ch Lagr: second cumulant for L>2} satisfy this property;
in fact, considering a certain quark mass, say $m_i$, tending to zero, we have:
\begin{equation}\label{Eff ch Lagr: chiral limit of chi and c4}
\chi \simeq \frac{F_\pi B_m m_i}{2} ~,~~~
c_4 \simeq -\frac{F_\pi B_m m_i}{2} ~,~~~{\rm for}~~ m_i \to 0 .
\end{equation}
Or, also, if we take $m_1 = \ldots = m_L \equiv m$, we find that:
\begin{equation}\label{Eff ch Lagr: chiral limit of chi and c4 - bis}
\chi \simeq \frac{F_\pi B_m m}{2L} ~,~~~
c_4 \simeq -\frac{F_\pi B_m m}{2L^3} ~,~~~{\rm for}~~ m \to 0 .
\end{equation}
The result found for the topological susceptibility $\chi$ in this limit is in agreement with what predicted by the relevant (flavour-singlet) Ward-Takahashi identities \cite{Crewther1977-1979}.

In the next sections, we shall consider different effective Lagrangian models which include the flavour-singlet meson field and also implement the $U(1)$ axial anomaly of the fundamental theory.
As we have said in the Introduction, in the last decades there were essentially two different ``schools of thought'' debating on how to address this issue: the first assumes that the dominant fluctuations are semiclassical instantons, while the second is based upon the large-$N_c$ limit of a $SU(N_c)$ gauge theory, and assumes that the dominant fluctuations are not semiclassical but quantum.
The model that we shall consider in Sec. 3 (the so-called \emph{Extended (Non-)Linear sigma model}) belongs to the first trend, while the model of Witten, Di Vecchia, Veneziano, \emph{et al.}, that we shall consider in Sec. 4, belongs to the second one.

\section{The ``Extended (Non-)Linear sigma model''}

The first effective Lagrangian model with the inclusion of the flavour-singlet meson field that we consider was originally proposed in Refs. \cite{ELSM1} to study the chiral dynamics at $T=0$, and later used in many different contexts (e.g., at non-zero temperature, around the chiral transition): in particular, 't Hooft (see Refs. \cite{ELSM2,ELSM3} and references therein) argued that it reproduces, in terms of an effective theory, the $U(1)$ axial breaking caused by instantons in the fundamental theory.
For brevity, from now on we shall refer to it as the \emph{Extended Linear sigma} ($EL_\sigma$) \emph{model}.
This model is described by the following Lagrangian:
\begin{equation}\label{'t Hooft effective Lagrangian}
\mathscr{L}(U,U^\dagger)= \mathscr{L}_0(U,U^\dagger) + \frac{B_m}{2\sqrt{2}}\Tr\left[\mathcal{M} U +\mathcal{M}^\dagger U^{\dagger}\right] + \mathscr{L}_I(U,U^\dagger) ,
\end{equation}
where $\mathscr{L}_0(U,U^\dagger)$ is the Lagrangian of the so-called \emph{Linear sigma model}, originally proposed in Ref. \cite{GL1960} but later elaborated on and extended:
\begin{equation}\label{Lagrangian of sigma model}
\begin{split}
\mathscr{L}_0(U,U^{\dagger})& = \frac{1}{2}\Tr [\partial_\mu U \partial^\mu U^{\dagger}] - V_0(U,U^{\dagger}) ,\\
V_0(U,U^{\dagger})&=\frac{1}{4}\lambda_\pi^2\Tr [( UU^{\dagger}-\rho_\pi \mathbf{I} )^2] + \frac{1}{4}\lambda_\pi^{'2}\left[\Tr(UU^\dagger)\right]^2 ,
\end{split}
\end{equation}
while $\mathscr{L}_I(U,U^\dagger)$ is the term which is claimed to describe, in terms of the effective variables, the $2L$-fermions interaction vertex generated by the instantons. Its form is:
\begin{equation}\label{instanton term in the Lagrangian}
\mathscr{L}_I(U,U^\dagger) = \kappa (\det U + \det U^\dagger ) ,
\end{equation}
where $\kappa$ is a constant which (according to 't Hooft) is expected to be proportional to the typical instanton factor $e^{-8\pi^2/g^2}$ \cite{tHooft1976}.
In this model, the mesonic effective fields are represented by a $L\times L$ complex matrix $U_{ij}$ which can be written, in terms of the quark fields, as:
\begin{equation}\label{Fundamental field U in terms of quark fields}
U_{ij}\sim \Overline[2]{q}_j \left(\frac{1+\gamma_5}{2}\right) q_i = \Overline[2]{q}_{jR}q_{iL} ,
\end{equation}
up to a multiplicative constant.
Under a chiral transformation \eqref{transformation SU(L)_L x SU(L)_R x U(1)_A}
the field $U$ transforms as:
\begin{equation}\label{chiral U transformation}
U \rightarrow \tilde{V}_L U \tilde{V}_R^\dagger ,
\end{equation}
and, as a consequence, the determinant of the field $U$ varies as:
\begin{equation}\label{chiral variation of detU}
\det U \rightarrow \det (\tilde{V}_L) \det (\tilde{V}_R)^* \det U .
\end{equation}
Therefore, the term \eqref{instanton term in the Lagrangian} is invariant under $SU(L)_L\otimes SU(L)_R \otimes U(1)_V$, while under a $U(1)_A$ transformation, $U \to e^{i2\alpha} U$, it varies as:
\begin{equation}\label{axial transformation of 't Hooft term}
\kappa (\det U + \det U^\dagger ) \rightarrow \kappa(e^{i2L\alpha} \det U + e^{-i2L\alpha} \det U^\dagger) .
\end{equation}
When using this model in our work, we have found more convenient to set the mass matrix in the real diagonal form $M=\diag (m_1,\ldots , m_L)$, by performing a $U(1)_A$ rotation of the field $U$ with $\alpha=-\frac{\theta}{2L}$, that is:
\begin{equation}\label{theta rotation of U to make the mass term real}
U\rightarrow e^{-i\theta /L}\, U .
\end{equation}
After this rotation, the Lagrangian \eqref{'t Hooft effective Lagrangian} is modified as:
\begin{equation}\label{'t Hooft effective Lagrangian after rotation}
\mathscr{L}(U,U^\dagger) = \mathscr{L}_0(U,U^\dagger) + \frac{B_m}{2\sqrt{2}}\Tr\left[M(U + U^{\dagger})\right] + \kappa(e^{-i\theta}\det U + e^{i\theta} \det U^\dagger ) .
\end{equation}
For what concerns the potential $V_0(U,U^\dagger)$ appearing in Eq.
\eqref{Lagrangian of sigma model}, we remind that the parameter $\rho_\pi$
is responsible for the fate of the chiral symmetry $SU(L)_L \otimes SU(L)_R$.
In particular, if (as it happens at $T=0$) $\rho_\pi>0$, then the vacuum
expectation value $\Overline[2]{U}$ of the mesonic field $U$ (i.e., the value
of $U$ for which the potential is at the minimum) is
(even in the chiral limit $M=0$) different from zero and of the form
 $\Overline[2]{U}|_{\rho_\pi>0} = v\,\mathbf{I}$,
meaning that the chiral symmetry is spontaneously broken down to the vectorial
$SU(L)_V$ subgroup.

If we are interested in describing only the low-energy dynamics of the effective pseudoscalar degrees of freedom (that is, the Goldstone [or \emph{would-be-}Goldstone] bosons), we can decouple the scalar massive fields by letting $\lambda_\pi^2\rightarrow \infty$: in fact, in this way, we are implementing the \emph{static limit} for the scalar fields, giving them infinite mass. In this limit, looking at the potential term in \eqref{Lagrangian of sigma model}, we are forcing the constraint $UU^\dagger=\rho_\pi \, \mathbf{I} \equiv\frac{F_\pi^2}{2} \, \mathbf{I}$, which implies $\Tr (UU^\dagger )=const.$: therefore, the term proportional to $\lambda_\pi^{'2}$ is just an irrelevant constant term, which can be dropped.
So, we shall neglect the scalar degrees of freedom and consider:
\begin{equation}\label{U form for T<Tc}
U=\frac{F_\pi}{\sqrt{2}}\, U' ~,~~~ U' \in U(L) .
\end{equation}
In this way, the Lagrangian of the model reduces to:
\begin{equation}\label{extended non-linear sigma model}
\mathscr{L}=\frac{1}{2}\Tr \left[\partial_\mu U \partial^\mu U^\dagger\right]
- V(U, U^\dagger) ,
\end{equation}
where the potential $V$ is (apart from a trivial constant):
\begin{equation}\label{ELsm: potential}
V(U,U^\dagger)=-\frac{B_m}{2\sqrt{2}}\Tr\left[M(U + U^{\dagger})\right] -\kappa(e^{-i\theta}\det U + e^{i\theta} \det U^\dagger ) .
\end{equation}
For brevity, from now on we shall refer to it as the \emph{Extended Non-Linear sigma} ($ENL_\sigma$) \emph{model}.
Setting $M$ in the usual diagonal form and $U$ as in \eqref{Diagonal form of U}
(but without the constraint \eqref{Constraint of the determinant of U} since
now $U'$ belongs to $U(L)$), we find:
\begin{equation}\label{ELsm: explicit potential}
V(\vec{\alpha})=-\frac{F_\pi B_m}{2}\sum\limits_{j=1}^L m_j \cos\alpha_j - 2\kappa\left(\frac{F_\pi}{\sqrt{2}}\right)^L \cos\left(\theta -\sum\limits_{j=1}^L \alpha_j\right) .
\end{equation}
The minimization equation is, therefore:
\begin{equation}\label{Minimization equation in ELsm}
\frac{\partial V(\vec{\alpha})}{\partial \alpha_i}= \frac{F_\pi B_m}{2}m_i \sin\alpha_i -2\kappa\left(\frac{F_\pi}{\sqrt{2}}\right)^L \sin\left(\theta -\sum\limits_{j=1}^L \alpha_j\right) = 0 .
\end{equation}
Again (as in the previous section), if we set $\theta=0$ the solution of the equation is $\alpha_j=0$ $\forall j$: we can thus consider both $\theta\ll 1$ and $\alpha_j\ll 1$ $\forall j$; moreover, from \eqref{ELsm: explicit potential} we see that the change $\theta \rightarrow -\theta$ is equivalent to the change $\alpha_j \rightarrow -\alpha_j$ $\forall j$. Therefore we can expand the phases $\alpha_j$ in powers of $\theta$, as in the previous section, but keeping only the odd-power terms. So, we set:
\begin{equation}\label{ELsm: expansion in theta of the phases alpha}
\alpha_i=A_i \theta + C_i \theta^3 +\ldots ,
\end{equation}
where the coefficients $A_i$ and $C_i$ have to be determined from the minimization condition. Inserting \eqref{ELsm: expansion in theta of the phases alpha} in \eqref{Minimization equation in ELsm} and expanding up to $\theta^3$, we have:
\begin{equation}\label{ELsm: expansion in theta of the explicit potential}
\begin{aligned}
\frac{\partial V(\vec{\alpha})}{\partial \alpha_i} & =\left[\frac{F_\pi B_m m_i}{2}A_i - 2\kappa\left(\frac{F_\pi}{\sqrt{2}}\right)^L\left(1-\sum_j A_j\right)\right] \theta \\
& + \left[\frac{F_\pi B_m m_i}{2} \left( C_i -\frac{1}{6}A_i^3 \right) + 2\kappa\left(\frac{F_\pi}{\sqrt{2}}\right)^L \sum_j C_j \right. \\
&\left. + 2\kappa \left(\frac{F_\pi}{\sqrt{2}}\right)^L \frac{1}{6}\left(1-\sum_j A_j\right)^3\right]\theta^3 +\ldots= 0 .
\end{aligned}
\end{equation}
Requiring that these equalities are satisfied order by order in $\theta$,
we derive the following expressions for the coefficients $A_i$ and $C_i$:
\begin{equation}\label{ELsm: linear order coefficient}
A_i= \frac{2\kappa\left(\frac{F_\pi}{\sqrt{2}}\right)^L}{\frac{F_\pi B_m \bar{m}}{2} + 2\kappa\left(\frac{F_\pi}{\sqrt{2}}\right)^L} \,\frac{\bar{m}}{m_i} ,
\end{equation}
\begin{equation}\label{ELsm: cubic order coefficient}
\begin{aligned}
C_i=& \frac{1}{6} \frac{2\kappa\left(\frac{F_\pi}{\sqrt{2}}\right)^L}{\left(\frac{F_\pi B_m \bar{m}}{2} + 2\kappa\left(\frac{F_\pi}{\sqrt{2}}\right)^L\right)^4} \,\frac{\bar{m}}{m_i} \\
\times & \left\{\frac{F_\pi B_m \bar{m}}{2}\left[\left(2\kappa\left(\frac{F_\pi}{\sqrt{2}}\right)^L\right)^2\left(\frac{\bar{m}}{m_i}\right)^2-\left(\frac{F_\pi B_m \bar{m}}{2}\right)^2\right] \right.\\ + & \left. \left(2\kappa\left(\frac{F_\pi}{\sqrt{2}}\right)^L\right)^3\left[\left(\frac{\bar{m}}{m_i}\right)^2-\sum_j\left(\frac{\bar{m}}{m_j}\right)^3\right] \right\} ,
\end{aligned}
\end{equation}
with $\bar{m}$ defined in Eq. \eqref{m-bar}.
Substituting the form \eqref{ELsm: expansion in theta of the phases alpha} in \eqref{ELsm: explicit potential} and expanding up to the order $\theta^4$, we find:
\begin{equation}\label{ELsm: theta dependence of the potential}
\begin{aligned}
V(\theta) & = const. + \frac{1}{2}\left[\frac{F_\pi B_m}{2}\sum_j m_j A_j^2 + 2\kappa\left(\frac{F_\pi}{\sqrt{2}}\right)^L (1-\sum_j A_j)^2 \right] \theta^2 \\
& + \frac{1}{24} \left[24\frac{F_\pi B_m}{2}\sum_j m_j A_j C_j - \frac{F_\pi B_m}{2}\sum_j m_j A_j^4 \right. \\
& - \left. 48\kappa\left(\frac{F_\pi}{\sqrt{2}}\right)^L (1-\sum_j A_j)\sum_j C_j - 2\kappa\left(\frac{F_\pi}{\sqrt{2}}\right)^L(1-\sum_j A_j)^4\right] \theta^4 + \ldots
\end{aligned}
\end{equation}
Finally, substituting the relations \eqref{ELsm: linear order coefficient} and \eqref{ELsm: cubic order coefficient} into \eqref{ELsm: theta dependence of the potential}, we can directly read, inside the square brackets, the expressions of the topological susceptibility and of the second cumulant. We report here the final results:
\begin{equation}\label{ELsm: topological susceptibility}
\chi=\frac{F_\pi B_m \bar{m}}{2} \frac{2\kappa\left(\frac{F_\pi}{\sqrt{2}}\right)^L}{\frac{F_\pi B_m \bar{m}}{2} + 2\kappa\left(\frac{F_\pi}{\sqrt{2}}\right)^L} ~,
\end{equation}
\begin{equation}\label{ELsm: second cumulant}
\begin{aligned}
c_4=&-\frac{F_\pi B_m \bar{m}}{2} \frac{2\kappa\left(\frac{F_\pi}{\sqrt{2}}\right)^L}{\left(\frac{F_\pi B_m \bar{m}}{2} + 2\kappa\left(\frac{F_\pi}{\sqrt{2}}\right)^L\right)^4} \\ &\times \left[\left(2\kappa\left(\frac{F_\pi}{\sqrt{2}}\right)^L\right)^3\sum_j\left(\frac{\bar{m}}{m_j}\right)^3 + \left(\frac{F_\pi B_m \bar{m}}{2}\right)^3\right] .
\end{aligned}
\end{equation}

\subsection{Considerations on the results}

First of all, we notice that, if we take the (formal) limit $\kappa\rightarrow\infty$, the expressions for the topological susceptibility and for the second cumulant obtained in the $ENL_\sigma$ model reduce precisely to those
found in the previous section using the Chiral Effective Lagrangian.
To explain this fact, it is sufficient to observe that the flavour-singlet squared mass takes a contribution from the term proportional to $\kappa$ in the
Lagrangian [see Eq. \eqref{instanton term in the Lagrangian}, which, using
$U = ({F_\pi}/{\sqrt{2}})U'$ with $U' = e^{i\sqrt{\frac{2}{L}}\frac{S_\pi}{F_\pi}} \tilde{U}'$, $\tilde{U}' \in SU(L)$, see Eq. \eqref{U form for T<Tc}, gives
$M_{S_\pi}^2 = \frac{2L}{F_\pi^2} 2\kappa \left(\frac{F_\pi}{\sqrt{2}}\right)^L$
in the chiral limit of zero quark masses\dots]
So, implementing the limit $\kappa\rightarrow\infty$, we are sending the flavour-singlet mass to infinity, decoupling it from the theory, which thus reduces to the Chiral Effective Lagrangian discussed in the previous section.

We also remark that (assuming that the parameter $\kappa$ is independent of
the quark masses or, at least, that it has a finite non-vanishing value in the
chiral limit) the expressions \eqref{ELsm: topological susceptibility} and
\eqref{ELsm: second cumulant} have the right behaviour
\eqref{Eff ch Lagr: chiral limit of chi and c4}, in the chiral limit
$m_i \to 0$, or \eqref{Eff ch Lagr: chiral limit of chi and c4 - bis},
in the chiral limit $m_1 = \ldots = m_L \equiv m \to 0$, as predicted
by the relevant (flavour-singlet) Ward-Takahashi identities \cite{Crewther1977-1979}.

If, on the contrary, we take the \emph{infinite quark-mass limit}, by sending all $m_j\rightarrow\infty$ (which results in $\bar{m}\rightarrow\infty$),\footnote{This limit is clearly a bit stretched since, from the beginning, we have based all the discussion on the existence of $L$ light quarks. Nevertheless, it is interesting to formally investigate the trend of the results also in this limit.}
we find that (assuming, again, that the parameter $\kappa$ is independent of
the quark masses or, at least, that it has a finite, non-divergent value in the
infinite quark-mass limit) the expressions \eqref{ELsm: topological susceptibility} and \eqref{ELsm: second cumulant} become:
\begin{equation}\label{ELsm: pure-gauge limit of chi and c4}
\chi \to 2\kappa\left(\frac{F_\pi}{\sqrt{2}}\right)^L ~,~~~
c_4 \to - 2\kappa\left(\frac{F_\pi}{\sqrt{2}}\right)^L .
\end{equation}
In this way, we are implementing the static limit for the quarks, so that the theory should reduce to a pure Yang-Mills one.
Indeed, the results \eqref{ELsm: pure-gauge limit of chi and c4} are in agreement with the $\theta$ dependence of the vacuum energy density expected in a pure-gauge theory as derived in an \emph{instanton-gas model} \cite{CDG1978}.
In fact, in this case one finds that:
\begin{equation}\label{theta dependence of vacuum energy in presence of instantons}
\epsilon_{vac}(\theta)\simeq \: const. - K \cos\theta
= const. + \frac{1}{2}K\theta^2 - \frac{1}{24}K\theta^4 + \ldots ,
\end{equation}
that, by virtue of Eq. \eqref{susceptibility and c4 in vacuum energy density},
leads to the relation $\chi = - c_4 = K$, which, taking
$K = 2\kappa\left(\frac{F_\pi}{\sqrt{2}}\right)^L$, is satisfied by the
results \eqref{ELsm: pure-gauge limit of chi and c4}.

\section{The effective Lagrangian model of Witten, Di Vecchia, Veneziano, \emph{et al.}}

A different chiral effective Lagrangian, with the inclusion of the flavour-singlet meson field, which implements the $U(1)$ axial anomaly of the fundamental theory, was proposed by Witten, Di Vecchia, Veneziano, \emph{et al.} \cite{WDV1,WDV2,WDV3}: for brevity, in the following we shall refer to this model as the \emph{WDV model}. Even if this model was derived and fully justified in the framework of the $1/N_c$ expansion (i.e., in the limit $N_c\rightarrow\infty$), the numerical results obtained using the $WDV$ model with $N_c=3$ are quite consistent with the real-world (experimental) values. This model is described by the Lagrangian (see Ref. \cite{WDV2} for a complete derivation):
\begin{equation}\label{Witten effective Lagrangian with Q}
\begin{split}
\mathscr{L}(U,U^\dagger,Q) &= \mathscr{L}_0(U,U^\dagger)
+ \frac{B_m}{2\sqrt{2}}\Tr[M(U +U^{\dagger})] \\
& + \frac{i}{2}Q(x) \Tr\left[\log U -\log U^\dagger\right]
+ \frac{1}{2A}Q^2(x) + \theta Q(x) ,
\end{split}
\end{equation}
where $\mathscr{L}_0(U,U^\dagger)$ is the Lagrangian of the
\emph{Linear sigma model}, reported in Eq. \eqref{Lagrangian of sigma model};
$Q(x)$ is the topological charge density and is introduced here as an \emph{auxiliary} field, while $A$ is a parameter which (at least in the large-$N_c$ limit) can be identified with the topological susceptibility in the pure Yang-Mills theory
($A = -i \int d^4 x \langle T Q(x) Q(0) \rangle\vert_{YM}$).
One immediately sees that the ``anomalous'' term $\mathscr{L}_{anom} \equiv \frac{i}{2}Q(x) \Tr\left[\log U -\log U^\dagger\right]$ in Eq. \eqref{Witten effective Lagrangian with Q} is invariant under $SU(L)_L\otimes SU(L)_R \otimes U(1)_V$, while under a $U(1)_A$ transformation, $U \to e^{i2\alpha} U$, it transforms as:
\begin{equation}\label{effective Lagrangian transformation under axial U(1)}
\mathscr{L}_{anom} \rightarrow \mathscr{L}_{anom} - 2L\alpha Q ,
\end{equation}
so correctly reproducing the $U(1)$ axial anomaly of the fundamental theory.\footnote{We recall here the criticism by Crewther (see also the third Ref. \cite{Crewther1977-1979}), Witten \cite{WDV1}, Di Vecchia and Veneziano \cite{WDV2} to the ``anomalous'' term \eqref{instanton term in the Lagrangian} of the $EL_\sigma$ model, which apparently does not correctly reproduce the $U(1)$ axial anomaly of the fundamental theory and, moreover, is inconsistent with the $1/N_c$ expansion.} 

According to what one is investigating, it may be convenient to integrate out the auxiliary field $Q(x)$ using its equation of motion, i.e.,
\begin{equation}\label{Q equation of motion}
Q(x) = -A\left[\theta + \frac{i}{2}\Tr\left(\log U - \log U^\dagger\right)\right] .
\end{equation}
After the substitution, we are left with:
\begin{equation}\label{Witten effective Lagrangian after having integrated out Q}
\mathscr{L}(U,U^\dagger) = \mathscr{L}_0(U,U^\dagger)
+ \frac{B_m}{2\sqrt{2}}\Tr[M(U +U^{\dagger})]
- \frac{A}{2}\left[\theta + \frac{i}{2}\Tr(\log U - \log U^\dagger)\right]^2 .
\end{equation}
As we have done in the previous section for the $EL_\sigma$ model, we shall neglect the scalar degrees of freedom (retaining only the low-energy dynamics of the effective pseudoscalar degrees of freedom),
by taking the formal limit $\lambda_\pi^2\rightarrow \infty$ (i.e., by taking
the limit of infinite mass for the scalar fields), which, as we have shown,
implies the constraint \eqref{U form for T<Tc} for the matrix field $U$.
In this way, the Lagrangian of the model reduces to:
\begin{equation}\label{WDV model}
\mathscr{L}=\frac{1}{2}\Tr \left[\partial_\mu U \partial^\mu U^\dagger\right]
- V(U, U^\dagger) ,
\end{equation}
where the potential $V$ is (apart from a trivial constant):
\begin{equation}\label{WDV: potential}
V(U, U^\dagger) = -\frac{B_m}{2\sqrt{2}}\Tr\left[M(U+U^\dagger)\right]
+ \frac{A}{2}\left[\theta + \frac{i}{2}\Tr (\log U - \log U^\dagger )\right]^2 .
\end{equation}
Setting $M$ in the usual diagonal form and $U$ as in \eqref{Diagonal form of U}
(but without the constraint \eqref{Constraint of the determinant of U}),
we find the following expression for the potential:
\begin{equation}\label{WDV: explicit potential}
V(\vec{\alpha})=-\frac{F_\pi B_m}{2}\sum\limits_{j=1}^L m_j \cos\alpha_j + \frac{A}{2}\left(\theta - \sum\limits_{j=1}^L \alpha_j\right)^2 .
\end{equation}
Therefore, the minimization equation is:
\begin{equation}\label{Minimization equation in WDV}
\frac{\partial V(\vec{\alpha})}{\partial \alpha_i} = \frac{F_\pi B_m}{2} m_i \sin \alpha_i - A\left(\theta - \sum_j \alpha_j\right) = 0 .
\end{equation}
As usual, since we are interested in the limit of small $\theta$ and, therefore,
also of small phases $\alpha_i$ (in fact, $\theta=0$ implies that $\alpha_i=0$
$\forall i$), we can Taylor-expand the sine in Eq. \eqref{Minimization equation in WDV} up to the third order in the phases:
\begin{equation}\label{WDV: potential up to the third order in the fields}
\frac{\partial V(\vec{\alpha})}{\partial \alpha_i} \simeq \frac{F_\pi B_m}{2} m_i \left(\alpha_i - \frac{\alpha_i^3}{6} + \ldots \right) - A\left(\theta - \sum_j \alpha_j\right)=0 ,
\end{equation}
and, moreover, observing that in \eqref{WDV: explicit potential} the change $\theta\rightarrow -\theta$ corresponds to the change $\alpha_j \rightarrow -\alpha_j$ $\forall j$, we can use for each phase $\alpha_i$ the following expansion in $\theta$:
\begin{equation}\label{WDV: guess for the alphas}
\alpha_i = A_i \theta + C_i \theta^3 + \ldots
\end{equation}
Inserting the expressions \eqref{WDV: guess for the alphas} into Eq.
\eqref{WDV: potential up to the third order in the fields}, we find that:
\begin{equation}\label{WDV: expansion in theta of the explicit potential}
\begin{aligned}
\frac{\partial V(\vec{\alpha})}{\partial \alpha_i}(\theta) & =\left[\frac{F_\pi B_m m_i}{2}A_i - A \left(1-\sum_j A_j\right)\right] \theta \\
& + \left[\frac{F_\pi B_m m_i}{2} \left(C_i - \frac{1}{6} A_i^3\right) + A \sum_j C_j \right]\theta^3 +\ldots= 0 .
\end{aligned}
\end{equation}
Requiring that these equalities are satisfied order by order in $\theta$, we find the following expressions for the coefficients $A_i$ and $C_i$:
\begin{equation}\label{WDV: linear order coefficient of the alphas}
A_i =  \frac{A}{\frac{F_\pi B_m \bar{m}}{2} + A} \frac{\bar{m}}{m_i} ,
\end{equation}
\begin{equation}\label{WDV: cubic order coefficient of the alphas}
C_i = \frac{1}{6}\left(\frac{A}{\frac{F_\pi B_m \bar{m}}{2} + A}\right)^3 \frac{\bar{m}}{m_i} \left[ \left(\frac{\bar{m}}{m_i}\right)^2 - \frac{A}{\frac{F_\pi B_m \bar{m}}{2} + A} \sum_j\left(\frac{\bar{m}}{m_j}\right)^3 \right] ,
\end{equation}
with $\bar{m}$ defined in Eq. \eqref{m-bar}.
Finally, Taylor-expanding the potential \eqref{WDV: explicit potential} up to the fourth order in the phases,
\begin{equation}\label{WDV: Taylor expansion of the potential up to the fourth order in the alphas}
V(\vec{\alpha})\simeq const. + \frac{F_\pi B_m}{4}\sum\limits_{j=1}^L m_j\left(\bar{\alpha}_j^2 - \frac{\bar{\alpha}_j^4}{12} + \ldots \right) + \frac{A}{2}\left(\theta - \sum\limits_{j=1}^L \bar{\alpha}_j\right)^2 ,
\end{equation}
and inserting the form \eqref{WDV: guess for the alphas}, with the expressions
\eqref{WDV: linear order coefficient of the alphas} and
\eqref{WDV: cubic order coefficient of the alphas} for the coefficients
$A_i$ and $C_i$ into Eq. \eqref{WDV: Taylor expansion of the potential up to the fourth order in the alphas}, we find:
\begin{equation}\label{WDV: theta dependence of the potential}
\begin{aligned}
V(\theta)=& \: const. + \frac{1}{2}\chi \theta^2 +
\frac{1}{24} c_4 \theta^4 + \ldots ,
\end{aligned}
\end{equation}
with the following expressions for the topological susceptibility $\chi$
and the second cumulant $c_4$ in this model:
\begin{equation}\label{WDV: topological susceptibility}
\chi=\frac{F_\pi B_m \bar{m}}{2}\frac{A}{\frac{F_\pi B_m \bar{m}}{2} + A} ,
\end{equation}
\begin{equation}\label{WDV: second cumulant}
c_4 = -\frac{F_\pi B_m \bar{m}}{2}\left(\frac{A}{\frac{F_\pi B_m \bar{m}}{2} + A}\right)^4 \sum\limits_{j=1}^L \left(\frac{\bar{m}}{m_j}\right)^3 .
\end{equation}

\subsection{Considerations on the results}

At first, we notice that the result \eqref{WDV: topological susceptibility} was already known in the literature \cite{WDV2}, but it was obtained by studying the two-point correlation function of the topological charge density operator $Q(x)$ rather than by means of the $\theta$ expansion of the vacuum energy density; instead, for what concerns the result \eqref{WDV: second cumulant}, it has been derived for the first time in this paper.
If we consider the (formal) limit $A\rightarrow\infty$, the results \eqref{WDV: topological susceptibility}-\eqref{WDV: second cumulant} obtained in the $WDV$ model precisely reduce to those found in the framework of the Chiral Effective Lagrangian in Sec. 2.
The reason is similar to the one discussed in the previous section for the $ENL_\sigma$ model: being the anomalous term proportional to $A$ in the Lagrangian \eqref{WDV model}-\eqref{WDV: potential} quadratic in the flavour-singlet field
[using $U = ({F_\pi}/{\sqrt{2}})U'$ with $U' = e^{i\sqrt{\frac{2}{L}}\frac{S_\pi}{F_\pi}} \tilde{U}'$, $\tilde{U}' \in SU(L)$, see Eq. \eqref{U form for T<Tc}, it gives $M_{S_\pi}^2 = \frac{2LA}{F_\pi^2}$ in the chiral limit of zero quark masses \ldots], such limit corresponds to send the flavour-singlet mass to infinity, decoupling it from the theory, which thus reduces to the SU(L) Chiral Effective Lagrangian discussed in Sec. 2.

For what concerns the topological susceptibility, we also observe that the result \eqref{WDV: topological susceptibility} coincides with the result \eqref{ELsm: topological susceptibility} found in the $ENL_\sigma$ model provided that the following substitution is implemented:
\begin{equation}\label{Substitution to go from WDV to ELsm and vice versa}
A \longleftrightarrow 2\kappa\left(\frac{F_\pi}{\sqrt{2}}\right)^L .
\end{equation}
And this correspondence also applies to the expression for the flavour-singlet squared mass $M_{S_\pi}^2$.
Remarkably, this is not so for the second cumulant: indeed, even after such substitution, the result \eqref{WDV: second cumulant} does not turn into \eqref{ELsm: second cumulant}. This is due to the difference between the anomalous terms in Eqs. \eqref{WDV: potential}-\eqref{WDV: explicit potential} and \eqref{ELsm: potential}-\eqref{ELsm: explicit potential}:
while the anomalous term in Eqs. \eqref{WDV: potential}-\eqref{WDV: explicit potential} is purely quadratic in the combination $\theta - \frac{\sqrt{2L}}{F_\pi}S_\pi$ (or: $\theta - \sum_j \alpha_j$), the anomalous term in Eqs.
\eqref{ELsm: potential}-\eqref{ELsm: explicit potential} is the cosine of such a combination.

We also remark that the expressions \eqref{WDV: topological susceptibility} and
\eqref{WDV: second cumulant} have the right behaviour
\eqref{Eff ch Lagr: chiral limit of chi and c4}, in the chiral limit
$m_i \to 0$, or \eqref{Eff ch Lagr: chiral limit of chi and c4 - bis},
in the chiral limit $m_1 = \ldots = m_L \equiv m \to 0$, as predicted
by the relevant (flavour-singlet) Ward-Takahashi identities \cite{Crewther1977-1979}.

Instead, if we take the infinite quark-mass limit, by sending all $m_j\rightarrow\infty$ (which results in $\bar{m}\rightarrow\infty$), we find that:
\begin{equation}\label{WDV: pure-gauge limit of chi and c4}
\chi \to A ~,~~~ c_4 \to 0 .
\end{equation}
As we have already observed in the previous section, this limit is meant to ``freeze'' the dynamics of the quarks, reducing the model to a pure Yang-Mills one. So, we expect that in this limit the topological susceptibility coincides with that of the pure-gauge theory: it is exactly what happens in our case.
For what concerns the second cumulant, it is null in this infinite quark-mass limit. This is due to the fact that the $WDV$ model is built considering only the leading terms in the expansion in $1/N_c$ and, so, while it contains the term
$\frac{1}{2A}Q^2$ [see Eq. \eqref{Witten effective Lagrangian with Q}],
it does not contain also a term proportional to $Q^4$, which would contribute to the pure-gauge value of the second cumulant $c_4$:
indeed, this kind of term is of the next-to-leading order in $1/N_c$ (for a detailed discussion on the next-to-leading terms, see Ref. \cite{DNPV1981}).

\section{An ``Interpolating model'' with the inclusion of a U(1) axial condensate}

In this section, we shall consider another effective Lagrangian model (which
was originally proposed in Refs. \cite{EM1994} and elaborated on in Refs.
\cite{MM2003,EM2011,MM2013}), which is in a sense in-between the $EL_\sigma$ model and the $WDV$ model: for this reason we shall call it the \emph{Interpolating model}.
Indeed, in this model the $U(1)$ axial anomaly is implemented, as in the $WDV$ model \eqref{Witten effective Lagrangian with Q}, by properly introducing the auxiliary field $Q$, so that it correctly satisfies the transformation property \eqref{effective Lagrangian transformation under axial U(1)} under the chiral group.
Moreover, it also includes an interaction term proportional to the determinant of the mesonic field $U$, which is similar to the interaction term \eqref{instanton term in the Lagrangian} in the $EL_\sigma$ model, assuming that there is another $U(1)_A$-breaking condensate (in addition to the usual quark-antiquark chiral condensate $\langle \bar{q}q \rangle$).
This extra $U(1)$ chiral condensate has the form
$C_{U(1)} = \langle {\cal O}_{U(1)} \rangle$,
where, for a theory with $L$ light quark flavors, ${\cal O}_{U(1)}$ is a
$2L$-quark local operator that has the chiral transformation properties of
\cite{tHooft1976,KM1970,Kunihiro2009}
${\cal O}_{U(1)} \sim \displaystyle{{\det_{st}}(\bar{q}_{sR}q_{tL})
+ {\det_{st}}(\bar{q}_{sL}q_{tR}) }$,
where $s,t = 1, \dots ,L$ are flavor indices. The color indices (not
explicitly indicated) are arranged in such a way that
(i) ${\cal O}_{U(1)}$ is a color singlet, and (ii)
$C_{U(1)} = \langle {\cal O}_{U(1)} \rangle$ is a \emph{genuine} $2L$-quark
condensate, i.e., it has no \emph{disconnected} part proportional to some
power of the quark-antiquark chiral condensate $\langle \bar{q} q \rangle$;
the explicit form of the condensate for the cases $L=2$ and $L=3$ is
discussed in detail in the Appendix A of Ref. \cite{EM2011} (see also Ref. \cite{DM1995}).

The effective Lagrangian of the \emph{Interpolating} model is written in terms
of the topological charge density $Q$, the mesonic field
$U_{ij} \sim \bar{q}_{jR} q_{iL}$ (up to a multiplicative constant),
and the new field variable $X \sim {\det} \left( \bar{q}_{sR} q_{tL} \right)$
(up to a multiplicative constant), associated with the $U(1)$ axial
condensate:
\begin{equation}\label{Interpolating model Lagrangian with Q}
\begin{split}
\mathscr{L}(U,U^\dagger , X,& X^\dagger , Q)
=\frac{1}{2}\Tr [\partial_\mu U \partial^\mu U^{\dagger}]
+ \frac{1}{2}\partial_\mu X \partial^\mu X^{\dagger} \\
&-V_0(U,U^\dagger , X,X^\dagger)
+ \frac{i}{2}\omega_1 Q(x) \Tr\left[\log U -\log U^\dagger\right] \\
&+ \frac{i}{2}(1-\omega_1)Q(x)\left[\log X -\log X^\dagger\right]
+ \frac{1}{2A}Q^2(x) +\theta Q(x) ,
\end{split}
\end{equation}
where
\begin{equation}\label{potential of the interpolating model}
\begin{split}
V_0(U,U^\dagger , X,X^\dagger) &= \frac{1}{4}\lambda_\pi^2\Tr [( UU^{\dagger}-\rho_\pi \mathbf{I} )^2] + \frac{1}{4}\lambda_\pi^{'2}\left[\Tr(UU^\dagger)\right]^2 \\ & +\frac{1}{4}\lambda_X^2 [XX^\dagger - \rho_X]^2 - \frac{B_m}{2\sqrt{2}}\Tr[M(U + U^{\dagger})] \\& - \frac{\kappa_1}{2\sqrt{2}}[X^\dagger \det U + X\det U^\dagger ] . 
\end{split}
\end{equation}
Since under a chiral $U(L)_L\otimes U(L)_R$ transformation \eqref{transformation SU(L)_L x SU(L)_R x U(1)_A} the field $X$ transforms exactly as $\det U$ [see Eq. \eqref{chiral variation of detU}], i.e.,
\begin{equation}\label{trasfX}
X \rightarrow \det (\tilde{V}_L) \det (\tilde{V}_R)^* X ,
\end{equation}
[i.e., $X$ is invariant under $SU(L)_L\otimes SU(L)_R\otimes U(1)_V$,
while, under a $U(1)$ axial transformation, $X\rightarrow e^{i2L\alpha}X$],
we have that, in the chiral limit $M=0$, the effective Lagrangian
\eqref{Interpolating model Lagrangian with Q}
is invariant under $SU(L)_L\otimes SU(L)_R \otimes U(1)_V$,
while under a $U(1)$ axial transformation, it correctly transforms as
in Eq. \eqref{effective Lagrangian transformation under axial U(1)}.

As in the case of the $WDV$ model, the auxiliary field $Q(x)$ in \eqref{Interpolating model Lagrangian with Q} can be integrated out using its equation of motion:
\begin{equation}
Q(x)=-A\left\{\theta + \frac{i}{2}\left[\omega_1 \Tr (\log U - \log U^\dagger )+ (1-\omega_1)(\log X - \log X^\dagger) \right]\right\} .
\end{equation}
After the substitution, we obtain:
\begin{equation}\label{Interpolating model Lagrangian without Q}
\mathscr{L}(U,U^\dagger , X,X^\dagger)=\frac{1}{2}\Tr [\partial_\mu U \partial^\mu U^{\dagger}] + \frac{1}{2}\partial_\mu X \partial^\mu X^{\dagger}
-\tilde{V}(U,U^\dagger , X,X^\dagger) ,
\end{equation}
where
\begin{equation}\label{potential of the interpolating model after having integrated out Q}
\begin{split}
\tilde{V}(U,U^\dagger &, X,X^\dagger)=V_0(U,U^\dagger , X,X^\dagger) \\
&+\frac{A}{2}\left\{\theta + \frac{i}{2}\left[\omega_1 \Tr (\log U - \log U^\dagger)+ (1-\omega_1)(\log X - \log X^\dagger )\right]\right\}^2 .
\end{split}
\end{equation}
Let us now briefly focus on the interaction term between $U$ and $X$ in Eqs.
\eqref{Interpolating model Lagrangian with Q}-\eqref{potential of the interpolating model}:
\begin{equation}\label{interaction term between X e U}
\mathscr{L}_{int}=\frac{\kappa_1}{2\sqrt{2}}[X^\dagger \det U + X\det U^\dagger ] .
\end{equation}
This term has a form very similar to the ``instantonic'' term \eqref{instanton term in the Lagrangian} of the $EL_\sigma$ model, but, differently from it, this term is invariant under the entire chiral group $U(L)_L \otimes U(L)_R$.\footnote{Assuming that the field $X$ has a non-zero vacuum expectation value $\Overline[2]{X}$ (which is the case if the parameter $\rho_X$ in the potential \eqref{potential of the interpolating model} is positive: see also Eq. \eqref{Parametrization of the field X} below\dots) and expanding $\det U = (F_\pi/\sqrt{2})^L e^{i\sqrt{2L}{S_\pi}/{F_\pi}}$ and $X=\Overline[2]{X}e^{i{S_X}/{\Overline[2]{X}}}$ in powers of the (pseudoscalar) excitations $S_\pi$ and $S_X$, one finds that $\mathscr{L}_{int}$ is quadratic at the leading order in the fields: considering for simplicity the chiral limit $M=0$ (and $\theta=0$), this term and the ``anomalous'' term (the last term in Eqs. \eqref{potential of the interpolating model after having integrated out Q} and \eqref{Interpolating: potential}) generate a squared-mass matrix for the fields $S_\pi$ and $S_X$, whose eigenstates are two different non-zero-mass singlets, called $\eta'$ and $\eta_X$ (see the original Refs. \cite{EM1994,MM2003,EM2011} for more details). This is what happens at $T=0$. Instead, at non-zero temperature, above the chiral transition, where $\Overline[2]{U}=0$ (and $U$ is thus ``linearized''), assuming that $\Overline[2]{X}$ is still different from zero (and, moreover, $\omega_1=0$; see Ref. \cite{MM2013}), one finds that, expanding in the fields:
\begin{equation}\label{Instantonic term in the Interpolating model}
\mathscr{L}_{int} = \kappa [\det U + \det U^\dagger ] + \ldots ~,~~~ \text{with:}~~ \kappa\equiv \frac{\kappa_1\Overline[2]{X}}{2\sqrt{2}} .
\end{equation}
In this case, therefore, the leading-order term in the fields has exactly the same form of the ``instantonic'' term \eqref{instanton term in the Lagrangian}: the dots in Eq. \eqref{Instantonic term in the Interpolating model} stay for
higher-order interaction terms containing also $S_X$.}

As usual, proceeding as we have done in the previous sections for the
$EL_\sigma$ model and the $WDV$ model, we shall neglect the
scalar degrees of freedom (retaining only the low-energy dynamics of the
effective pseudoscalar degrees of freedom), by taking the formal limits
$\lambda_\pi^2\rightarrow \infty$ and $\lambda_X^2\rightarrow \infty$
(i.e., by taking the limit of infinite mass for the scalar fields), which,
in addition to the constraint \eqref{U form for T<Tc} for the matrix field $U$,
also implies the analogous constraint
$XX^\dagger = \rho_X \equiv\frac{F_X^2}{2}$ for the $X$ field, i.e.,
\begin{equation}\label{Parametrization of the field X}
X=\frac{F_X}{\sqrt{2}}\, e^{i\beta} ,
\end{equation}
having introduced the decay constant $F_X$ of the field $X$, analogous to the
decay constant $F_\pi$ of the pions.
In this way, the Lagrangian of the model reduces to:
\begin{equation}\label{Interpolating model Lagrangian}
\mathscr{L}(U,U^\dagger , X,X^\dagger)=\frac{1}{2}\Tr [\partial_\mu U \partial^\mu U^{\dagger}] + \frac{1}{2}\partial_\mu X \partial^\mu X^{\dagger}
-V(U,U^\dagger , X,X^\dagger) ,
\end{equation}
where the potential $V$ is (apart from a trivial constant):
\begin{equation}\label{Interpolating: potential}
\begin{aligned}
V(U,U^\dagger,X,X^\dagger)&= - \frac{B_m}{2\sqrt{2}}\Tr[M(U+ U^{\dagger})] - \frac{\kappa_1}{2\sqrt{2}}[X^\dagger \det U + X \det U^\dagger ] \\
& +\frac{A}{2}\left\{\theta + \frac{i}{2}\left[\omega_1 \Tr (\log U - \log U^\dagger) + (1-\omega_1)(\log X - \log X^\dagger)\right]\right\}^2 .
\end{aligned}
\end{equation}
Setting $M$ in the usual diagonal form, $U$ as in Eq. \eqref{Diagonal form of U}
(but without the constraint \eqref{Constraint of the determinant of U}) and
the analogous parametrization \eqref{Parametrization of the field X}
for the field $X$, where the phase $\beta$ (exactly as the phases $\alpha_j$)
is constant with respect to $x$, we find the following expression for the
potential:
\begin{equation}\label{Interpolating: explicit potential}
\begin{aligned}
V(\vec{\alpha},\beta)= &-\frac{F_\pi B_m}{2}\sum\limits_{j=1}^L m_j \cos\alpha_j - c \cos\left(\beta-\sum\limits_{j=1}^L \alpha_j\right) \\
& +\frac{A}{2}\left[\omega_1\sum\limits_{j=1}^L \alpha_j + (1-\omega_1)\beta - \theta\right]^2 ,
\end{aligned}
\end{equation}
where we have defined:
\begin{equation}\label{Interpolating: definition of c}
c\equiv \kappa_1\frac{F_X}{2}\left(\frac{F_\pi}{\sqrt{2}}\right)^L .
\end{equation}
In order to find the minimum of the potential, we have to solve the following system of minimization equations:
\begin{equation}\label{Minimization equations in the interpolating model}
\frac{\partial V(\vec{\alpha}, \beta)}{\partial \alpha_i} = 0 \quad \forall i=1,\ldots,L ~,~~~
\frac{\partial V(\vec{\alpha}, \beta)}{\partial \beta} = 0 ,
\end{equation}
which, after a slight rearrangement, read as follows:
\begin{equation}\label{Minimization equations in the interpolating model rearranged}
\left\{
\begin{aligned}
&\frac{F_\pi B_m}{2}m_i\sin\alpha_i +A \left[\omega_1\sum\limits_{j=1}^L \alpha_j + (1-\omega_1)\beta - \theta\right]=0 ,\\
\\
&c\sin\left(\beta-\sum_j\alpha_j\right) + A(1-\omega_1) \left[\omega_1\sum\limits_{j=1}^L \alpha_j + (1-\omega_1)\beta - \theta\right]=0 .
\end{aligned}
\right.
\end{equation}
It is easy to check that, in the case $\theta=0$, setting $\beta=0$ and $\alpha_j=0$ $\forall j$ puts the potential in its minimum. So, if we consider the case
$\theta\ll 1$, we are allowed to use for the phases $\alpha_i$ and $\beta$
the following Taylor expansion in powers of $\theta$:
\begin{equation}\label{Interpolating: expansion in theta of all the phases}
\alpha_i=A_i \theta + B_i \theta^2 + C_i \theta^3 + \ldots ~,~~~
\beta = W \theta + Y \theta^2 + Z \theta^3 + \ldots
\end{equation}
The coefficients $A_i$, $B_i$, $C_i$, $W$, $Y$, $Z$ have to be determined by solving (order by order in $\theta$) the system \eqref{Minimization equations in the interpolating model rearranged}.
Looking at the equations \eqref{Minimization equations in the interpolating model rearranged}, it is easy to see that the change $\theta \rightarrow -\theta$ corresponds to the changes $\alpha_j \rightarrow -\alpha_j$ $\forall j$ and $\beta \rightarrow -\beta$, and, as a consequence, the coefficients of the even powers of $\theta$ in the expansions \eqref{Interpolating: expansion in theta of all the phases} must vanish:
\begin{equation}\label{Interpolating: quadratic coefficients of the expansions are equal to zero}
Y=0 ~,~~~ B_i=0 \quad \forall i .
\end{equation}
Concerning the coefficients of the odd powers of $\theta$, the following expressions are found:
\begin{equation}\label{Interpolating: linear coefficient of beta}
W=\frac{A \left[ 1+\frac{F_\pi B_m\bar{m}}{2c}(1-\omega_1) \right]}
{\frac{F_\pi B_m\bar{m}}{2}\left(1+\frac{A(1-\omega_1)^2}{c}\right)+A} ,
\end{equation}
\begin{equation}\label{Interpolating: linear coefficient of alphas}
A_i= \frac{A}{\frac{F_\pi B_m\bar{m}}{2}\left(1+\frac{A(1-\omega_1)^2}{c}\right)+A} \frac{\bar{m}}{m_i} ,
\end{equation}
\begin{equation}\label{Interpolating: cubic coefficient of beta}
\begin{aligned}
Z=&\frac{1}{6}\left(\frac{F_\pi B_m\bar{m}}{2}\right) A^3 \left[ \frac{F_\pi B_m\bar{m}}{2}\left(1+\frac{A(1-\omega_1)^2}{c}\right)+A \right]^{-4} \\
&\times \left[\left(\frac{F_\pi B_m\bar{m}}{2}\right)^2\frac{(1-\omega_1)^3}{c^3}\left(\frac{F_\pi B_m\bar{m}}{2} + A\omega_1 \right) \right. \\
& + \left. \left( 1 - \frac{A\omega_1(1-\omega_1)}{c} \right) \sum_j\left(\frac{\bar{m}}{m_j}\right)^3\right] ,
\end{aligned}
\end{equation}
and:
\begin{equation}\label{Interpolating: cubic coefficient of alphas}
\begin{aligned}
C_i = \frac{1}{6}& \left[ \frac{A}{\frac{F_\pi B_m\bar{m}}{2}\left(1+\frac{A(1-\omega_1)^2}{c}\right)+A} \right]^3 \frac{\bar{m}}{m_i} \\
\times & \left\{ \left(\frac{\bar{m}}{m_i}\right)^2 - \frac{A \left[\sum_j\left(\frac{\bar{m}}{m_j}\right)^3 + \left(\frac{F_\pi B_m\bar{m}}{2c}\right)^3 (1-\omega_1)^4 \right]}{\frac{F_\pi B_m\bar{m}}{2}\left(1+\frac{A(1-\omega_1)^2}{c}\right)+A} \right\} ,
\end{aligned}
\end{equation}
with $\bar{m}$ defined in Eq. \eqref{m-bar}.
Substituting the expressions \eqref{Interpolating: expansion in theta of all the phases} (with $B_i=Y=0$) into Eq. \eqref{Interpolating: explicit potential}
and expanding the potential up to the fourth order in $\theta$, we find:
\begin{equation}\label{Interpolating: potential up to the fourth order in theta}
\begin{aligned}
V(\theta)& \simeq const. {+} \frac{1}{2}\left\{\frac{F_\pi B_m}{2}\sum_j m_j A_j^2{+}c(W{-}\sum_j A_j)^2
{+} A\Big[\omega_1\sum_j A_j{+}(1{-}\omega_1)W{-}1\Big]^2\right\}\theta^2 \\
& {+} \frac{1}{24}\left\{24\,\frac{F_\pi B_m}{2}\sum_j m_j A_j C_j
{-} \frac{F_\pi B_m}{2}\sum_j m_j A_j^4{+}24\,c\,(W{-}\sum_j A_j)(Z{-}\sum_j C_j) \right. \\
& \left. {-} c\:(W{-}\sum_j A_j)^4 {+} 24A\Big[\omega_1\sum_j A_j{+}(1{-}\omega_1)W{-}1\Big] \Big[\omega_1\sum_j C_j{+}(1{-}\omega_1)Z\Big]\right\}\theta^4 + \ldots ,
\end{aligned}
\end{equation}
from which, after inserting the expressions \eqref{Interpolating: linear coefficient of beta}-\eqref{Interpolating: cubic coefficient of alphas}, we obtain the following expressions for the topological susceptibility $\chi$ and the second cumulant $c_4$ in this model:
\begin{equation}\label{Interpolating: topological susceptibility}
\chi=\frac{F_\pi B_m\bar{m}}{2} \:
\frac{A}{\frac{F_\pi B_m\bar{m}}{2}\left(1+\frac{A(1-\omega_1)^2}{c}\right)+A} ,
\end{equation}
\begin{equation}\label{Interpolating: second cumulant}
\begin{aligned}
c_4=-&\frac{F_\pi B_m\bar{m}}{2}\left[\frac{A}{\frac{F_\pi B_m\bar{m}}{2}\left(1+\frac{A(1-\omega_1)^2}{c}\right)+A}\right]^4 \\ & \times \left[\sum\limits_{j=1}^L\left(\frac{\bar{m}}{m_j}\right)^3+\left(\frac{F_\pi B_m\bar{m}}{2c}\right)^3(1-\omega_1)^4\right] .
\end{aligned}
\end{equation}

\subsection{Considerations on the results}

We first notice that the result \eqref{Interpolating: topological susceptibility} was originally found in Ref. \cite{EM1994}, but once again it was obtained by a different approach, i.e., by directly studying the two-point function of the field $Q(x)$. On the contrary, the result \eqref{Interpolating: second cumulant} has been derived in this paper for the first time.
Moreover, we notice that, if $\omega_1 \neq 1$, the topological susceptibility obtained in this \emph{Interpolating} model is smaller than the one obtained in the $WDV$ model, due to the positive (assuming $c>0$: see Refs. \cite{EM2011,MM2013}) corrective factor in the denominator.
If, instead, we set $\omega_1 = 1$ (which, as we shall comment in the next section, represents the most natural choice at $T=0$) the results for both $\chi$ and $c_4$ coincide precisely with those of the $WDV$ model (independently of the other parameters $\kappa_1$ and $F_X$ of the model). The explanation of this fact lies in the potential \eqref{Interpolating: explicit potential}: indeed, if we set $\omega_1=1$, we immediately see that, so as to obtain the minimum value for $V(\vec{\alpha},\beta)$, it is clear that we must set $\beta=\sum_j \alpha_j$, so that the cosine in the second term is equal to one. In this way, we find that the potential \eqref{Interpolating: explicit potential} coincides with the potential \eqref{WDV: explicit potential} of the $WDV$ model apart from a constant with respect to $\theta$: so, the final results for the topological susceptibility and for the second cumulant in the \emph{Interpolating} model with $\omega_1=1$ are indeed expected to coincide with those of the $WDV$ model.

\section{Conclusions: summary and analysis of the results}

In this conclusive section, we shall summarize the analytical results that we have found for the topological susceptibility $\chi$ and the second cumulant $c_4$ in the various cases that we have considered.
Moreover, we shall also report numerical estimates for these quantities, obtained both for $L=2$ and $L=3$ in the case of the Chiral Effective Lagrangian (see the discussion at the end of Sec. 2), and for $L=3$ in the other cases (effective Lagrangian models with the inclusion of the flavour-singlet meson field).\footnote{As discussed in detail in Ref. \cite{Veneziano1979},
when including the flavour-singlet meson field in the effective Lagrangian,
we must consider the case $L=3$, if we want to have a realistic description of
the physical world (at least at $T=0$): this is essentially due to
the fact that (see below) the value of $B m_s$, while being considerably larger
than $B m_u$ and $B m_d$, is comparable to (or even smaller than) the anomalous
contribution proportional to $2A/F_\pi^2$ in the meson squared mass matrix\dots}
For our numerical computations, the following values of the known parameters have been used:
\begin{itemize}
\item $A=\left(180\pm 5 \; \text{MeV}\right)^4$ (see Ref. \cite{DGP2005} and references therein).
\item $F_\pi=92.2 \pm 0.2$ MeV (see Ref. \cite{PDG2016-2017}, where the value of $f_\pi = \sqrt{2}F_\pi$ is reported).
\item For what concerns the parameter $B_m$, we shall rewrite it making use of the relation \eqref{B_m definition} in terms of the quantity $B$, which directly relates the quark masses to the light pseudoscalar meson masses. In particular, the following relations hold, at the leading order in the chiral perturbation theory:
\begin{equation}\label{formulae for B m_i}
\begin{aligned}
B m_u &= M_{\pi^0}^2 - \frac{1}{2}\left(M_{K^0}^2-M_{K^+}^2 + M_{\pi^+}^2\right) , \\
B m_d &= \frac{1}{2}\left(M_{K^0}^2-M_{K^+}^2 + M_{\pi^+}^2\right) , \\
B m_s &= \frac{1}{2}\left(M_{K^0}^2+M_{K^+}^2 - M_{\pi^+}^2\right) . \\
\end{aligned}
\end{equation}
So, these expressions can be numerically evaluated using the known values for the masses of the mesons $\pi^+$, $\pi^0$, $K^+$, $K^0$ \cite{PDG2016-2017}:
\begin{equation}\label{Meson masses values}
\begin{aligned}
& M_{\pi^+} = 139.57061(24) \; \text{MeV} , \\
& M_{\pi^0} = 134.9770(5) \; \text{MeV} , \\
& M_{K^+} = 493.677(16) \; \text{MeV} , \\
& M_{K^0} = 497.611(13) \; \text{MeV} .
\end{aligned}
\end{equation}
\item For what concerns the quantity $\kappa$, its value is not known \emph{a priori}. A possible way to evaluate it numerically is to make use of the relation among $\kappa$, $F_\pi$ and the meson masses, obtained within the $ENL_\sigma$ model in the case $L=3$:
\begin{equation}\label{relation for kappa}
M^2_{\eta'} + M^2_\eta - M_{K^0}^2 - M_{K^+}^2 = 6\kappa \frac{F_\pi}{\sqrt{2}}.
\end{equation}
Substituting the experimental values of the meson masses (in addition to those given in \eqref{Meson masses values} we need $M_\eta=547.862 \pm 0.017$ MeV and $M_{\eta'}=957.78 \pm 0.06$ MeV \cite{PDG2016-2017}), we find for this parameter the value $\kappa=1856.38\pm 4.04$ MeV.
\end{itemize}
The values of all the parameters we listed above allow us to evaluate numerically all the results coming from the Chiral Effective Lagrangian at the leading order $\mathcal{O}(p^2)$, the $ENL_\sigma$ model and the $WDV$ model. The situation of the \emph{Interpolating} model is more complicated: due to the fact that very little is known about its peculiar parameters, it is not possible to give a complete numerical form to the results found in this model. In particular:
\begin{itemize}
\item For what concerns the parameter $F_X$, only an upper bound is known
for it \cite{EM1994,MM2003,EM2011}: $\left|F_X\right| \leq 20$ MeV. 
\item For what concerns the parameter $\kappa_1$ (which was named ``$c_1$'' in the original papers), we cannot say too much, apart from the fact that (assuming $F_X \neq 0$) it cannot be zero (see Ref. \cite{EM2011} for a detailed discussion on the role of this parameter).
\item At last, concerning the parameter $\omega_1$, we observe that the Lagrangian of the $WDV$ model is obtained from that of the \emph{Interpolating} model by choosing $\omega_1=1$ (and then letting $F_X \to 0$). At low temperatures, one expects that the deviations from the $WDV$ Lagrangian are small, in some sense, and therefore that $\omega_1$ should not be much different from the unity near $T=0$ (on the other side, $\omega_1$ must necessarily be taken equal to zero above the chiral transition temperature, in order to avoid a singular behaviour of the anomalous term \cite{EM1994,MM2013}).
Therefore, $\omega_1=1$ seems to be the most natural choice for $T=0$: with this choice, all the numerical values coincide with those of the $WDV$ model, regardless of the values of the other (unknown) parameters of the model, i.e., $\kappa_1$ and $F_X$.
\end{itemize}
Here is (in the following two subsections) a summary of both analytical and numerical results. [We recall that $\bar{m}$ is defined in Eq. \eqref{m-bar}.]

\subsection{Topological susceptibility}

\begin{itemize}
\item \emph{Chiral Effective Lagrangian $\mathcal{O}(p^2)$}:
\begin{eqnarray}\label{Eff Ch Lagr: numerical topological susceptibility}
\chi =& \frac{F_\pi B_m\bar{m}}{2} \nonumber \\
\chi^{(L=2)} =& \left(77.25\pm 0.08 \; \text{MeV}\right)^4 \nonumber \\
\chi^{(L=3)} =& \left(76.91\pm 0.08 \; \text{MeV}\right)^4
\end{eqnarray}
\item \emph{ENL$_{\sigma}$ model}:
\begin{eqnarray}\label{ELsm: numerical topological susceptibility}
\chi =& \frac{F_\pi B_m \bar{m}}{2} \frac{2\kappa\left(\frac{F_\pi}{\sqrt{2}}\right)^L}{\frac{F_\pi B_m \bar{m}}{2} + 2\kappa\left(\frac{F_\pi}{\sqrt{2}}\right)^L} \nonumber \\
\chi^{(L=3)} =& \left(76.271\pm 0.085 \; \text{MeV}\right)^4
\end{eqnarray}
\item \emph{WDV model}:
\begin{eqnarray}\label{WDV: numerical topological susceptibility}
\chi =& \frac{F_\pi B_m \bar{m}}{2}\frac{A}{\frac{F_\pi B_m \bar{m}}{2} + A} \nonumber \\
\chi^{(L=3)} =& \left(76.283\pm 0.106 \; \text{MeV}\right)^4
\end{eqnarray}
\item \emph{Interpolating model}:
\begin{eqnarray}\label{IM: numerical topological susceptibility}
\chi =& \frac{F_\pi B_m\bar{m}}{2} \: \frac{A}{\frac{F_\pi B_m\bar{m}}{2}\left(1+\frac{A(1-\omega_1)^2}{c}\right)+A} \nonumber \\
\chi^{(L=3)}_{(\omega_1=1)} =& \left(76.283\pm 0.106 \; \text{MeV}\right)^4
\end{eqnarray}
\end{itemize}

\subsection{Second cumulant}

\begin{itemize}
\item \emph{Chiral Effective Lagrangian $\mathcal{O}(p^2)$}:
\begin{eqnarray}\label{Eff Ch Lagr: numerical second cumulant}
c_4 =& -\frac{F_\pi B_m\bar{m}}{2}\sum\limits_{j=1}^L\left(\frac{\bar{m}}{m_j}\right)^3 \nonumber \\
c_4^{(L=2)} =& -\left(11.05 \pm 0.49\right)\times 10^6 \; \text{MeV}^4 \nonumber \\
c_4^{(L=3)} =& -\left(10.30 \pm 0.46\right)\times 10^6 \; \text{MeV}^4
\end{eqnarray}
\item \emph{ENL$_{\sigma}$ model}:
\begin{eqnarray}\label{ELsm: numerical second cumulant}
c_4 =& -\frac{F_\pi B_m \bar{m}}{2} \frac{2\kappa\left(\frac{F_\pi}{\sqrt{2}}\right)^L}{\left(\frac{F_\pi B_m \bar{m}}{2} + 2\kappa\left(\frac{F_\pi}{\sqrt{2}}\right)^L\right)^4} \nonumber \\
&\times \left[\left(2\kappa\left(\frac{F_\pi}{\sqrt{2}}\right)^L\right)^3\sum_j\left(\frac{\bar{m}}{m_j}\right)^3 + \left(\frac{F_\pi B_m \bar{m}}{2}\right)^3\right] \nonumber \\
c_4^{(L=3)} =& -\left(9.007 \pm 0.426\right)\times 10^6 \; \text{MeV}^4
\end{eqnarray}
\item \emph{WDV model}:
\begin{eqnarray}\label{WDV: numerical second cumulant}
c_4 =& -\frac{F_\pi B_m \bar{m}}{2}\left(\frac{A}{\frac{F_\pi B_m \bar{m}}{2} + A}\right)^4 \sum\limits_{j=1}^L \left(\frac{\bar{m}}{m_j}\right)^3 \nonumber \\
c_4^{(L=3)} =& -\left(9.030 \pm 0.134\right)\times 10^6 \; \text{MeV}^4
\end{eqnarray}
\item \emph{Interpolating model}:
\begin{eqnarray}\label{IM: numerical second cumulant}
c_4 =& -\frac{F_\pi B_m\bar{m}}{2}\left(\frac{A}{\frac{F_\pi B_m\bar{m}}{2}\left(1+\frac{A(1-\omega_1)^2}{c}\right)+A}\right)^4 \nonumber \\
& \times \left[\sum\limits_{j=1}^L\left(\frac{\bar{m}}{m_j}\right)^3+\left(\frac{F_\pi B_m\bar{m}}{2c}\right)^3(1-\omega_1)^4\right] \nonumber \\
c_{4~(\omega_1=1)}^{(L=3)} =& -\left(9.030 \pm 0.134\right)\times 10^6 \; \text{MeV}^4
\end{eqnarray}
\end{itemize}
Let us make some remarks on these results.
We observe that, within the present accuracy, there are no significant numerical differences between the results found in the $ENL_\sigma$ model and those found in the $WDV$ model (or in the \emph{Interpolating} model with $\omega_1=1$),
even if the theoretical expressions for the topological susceptibility and the second cumulant are in principle different (even considering the correspondence \eqref{Substitution to go from WDV to ELsm and vice versa}: see the discussion in Sec. 4.1).
On the contrary, the numerical results found in the $ENL_\sigma$ model, the $WDV$ model, and the \emph{Interpolating} model with $\omega_1=1$, are sensibly different from those found using the Chiral Effective Lagrangian at order $\mathcal{O}(p^2)$.
In this respect, we must here recall that in Refs. \cite{MC2009,GM2015,GHVV2016} also the non-leading order (NLO) correction to the result for the topological susceptibility using the Chiral Effective Lagrangian have been computed, and it turned out that it is of the order of percent for physical quark masses. Starting from our results, we can derive the order of the corrections caused by the presence of the flavour singlet to the numerical values obtained using the Chiral Effective Lagrangian $\mathcal{O}(p^2)$, so as to make a comparison with that of the NLO corrections: for what concerns the topological susceptibility, these corrections are of the order of some percent and, so, are comparable with the NLO ones; for what concerns the second cumulant, instead, the corrections are considerably larger, being about the 12\%.

\subsection{Comparison of the results with the literature}

In the end, let us make a comparison between the above-reported numerical estimates and the available lattice results in the literature.
We first consider the topological susceptibility. The value of the topological susceptibility in \emph{full} QCD has been measured through Monte Carlo simulations on the lattice. We report here two recent results, obtained with $L=2+1$ light flavours with physical quark masses:
\begin{equation}\label{lattice results for topological susceptibility}
\begin{split}
\chi^{1/4} = 73(9) \; \text{MeV} ~\qquad \text{(see Ref. \cite{chi-lattice_1})};\\
\chi^{1/4} = 75.6(2.0) \; \text{MeV} \quad \text{(see Ref. \cite{chi-lattice_2})},
\end{split}
\end{equation}
where, for the second value, the error in parentheses has been obtained adding in quadrature the statistical error (1.8) and the systematic error (0.9).
These results are in perfect agreement (within the large errors) with all those found in our work.
In figure 1, the numerical values obtained for the topological susceptibility in our work are reported together with the lattice results.
\begin{figure}[!ht]
\begin{center}
\begin{minipage}[!ht]{8 cm}
\centering
\includegraphics[width=7.8 cm]{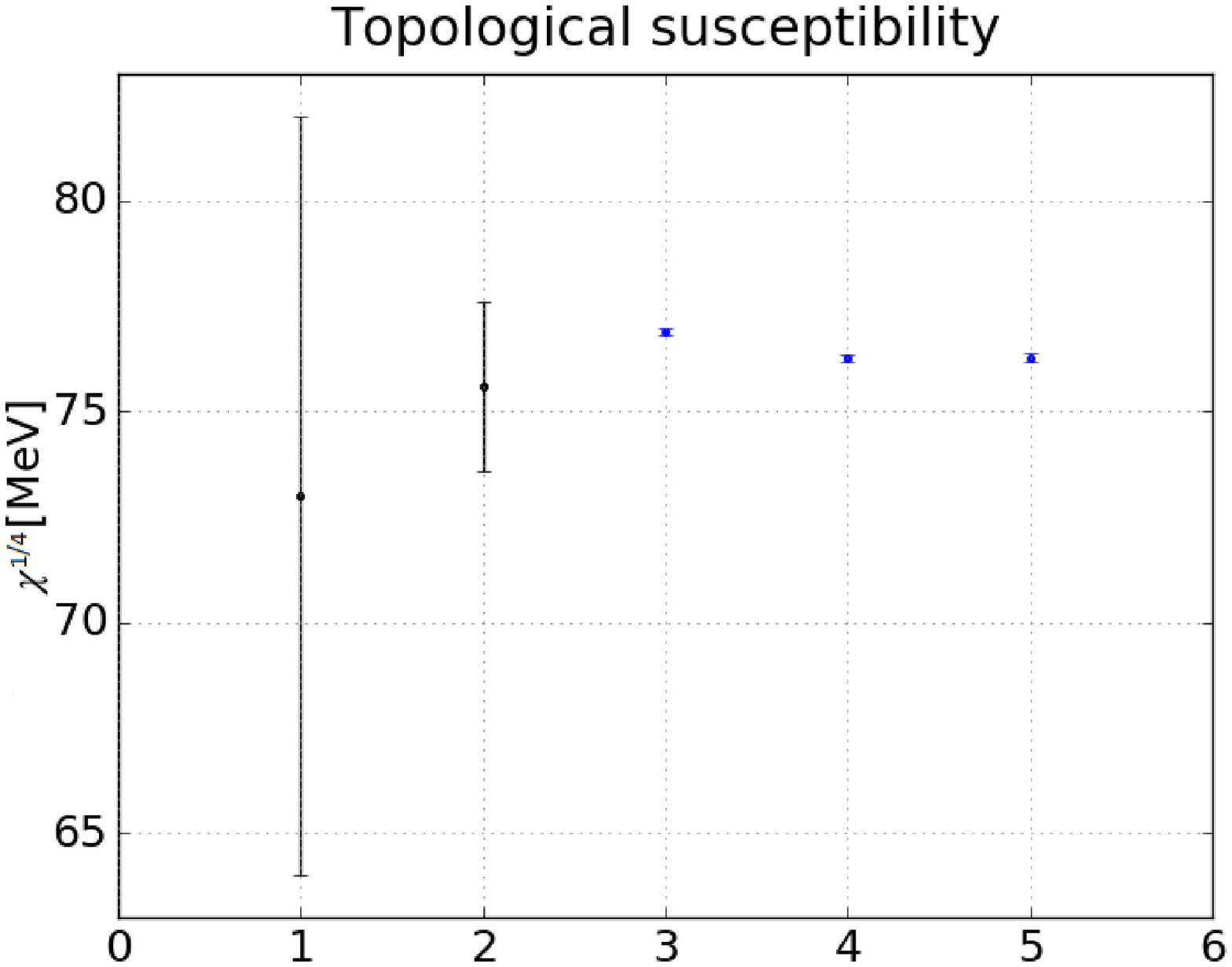}
\end{minipage}
\begin{minipage}[!ht]{7.2 cm}
\centering
\includegraphics[width=7.2 cm]{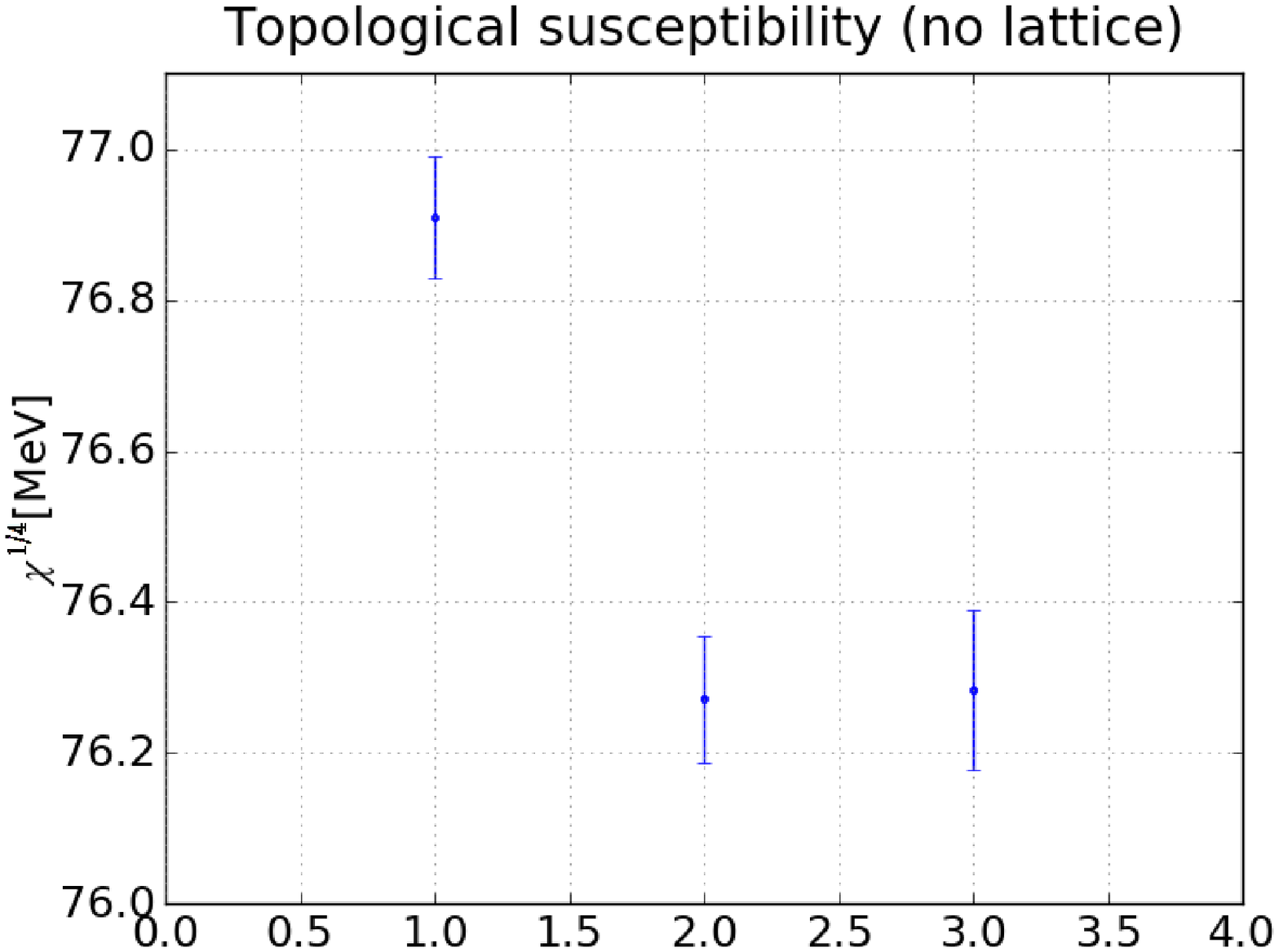}
\end{minipage}
\end{center}
\caption{On the left, the two lattice results
\eqref{lattice results for topological susceptibility} for the topological
susceptibility (in the \emph{full} theory with quarks) and the three
theoretical estimates for $L=3$, reported in Eqs.
\eqref{Eff Ch Lagr: numerical topological susceptibility},
\eqref{ELsm: numerical topological susceptibility}, and
\eqref{WDV: numerical topological susceptibility}-\eqref{IM: numerical topological susceptibility},
are shown (from left to right). On the right, only the three theoretical estimates are shown (in a different scale), so as to better compare them with each other.}
\end{figure}
With the help of this figure, we clearly see that the numerical value obtained using the Chiral Effective Lagrangian $\mathcal{O}(p^2)$ (the first point in the figure on the right) is clearly detached from the ones related to the $ENL_\sigma$ model and to the $WDV$ (or \emph{Interpolating}) model (respectively, the second and the third point in the figure on the right). Besides, these last two values are evidently compatible within the uncertainties.

Let us now move to the second cumulant. In lattice simulations, a quantity which is linked to the second cumulant is usually measured  rather than the second cumulant itself, due to a simpler definition on the lattice. We report here the definition of this quantity, usually called $b_2$ (a more detailed description of this parameter can be found in Ref. \cite{VP2009}):
\begin{equation}\label{b2 definition}
b_2 \equiv \frac{c_4}{12\chi} = -\frac{\langle Q_{tot}^4\rangle_{\theta=0}- 3 \langle Q_{tot}^2\rangle^2_{\theta=0}}{12 \langle Q_{tot}^2\rangle_{\theta=0}} .
\end{equation}
All the lattice determinations of this parameter at $T=0$ are obtained, to date, in $SU(N_c)$ pure-gauge frameworks, considering $N_c\geq3$: it must be taken into account that our final results have been obtained in a \emph{full} QCD framework. There are, in the literature, a number of results for $b_2$ at $N_c = 3$, obtained using different approaches (see Ref. \cite{VP2009} and references therein):
\begin{equation}\label{lattice results for b2 (1)}
\begin{aligned}
b_2&=-0.023(7) \quad \text{(cooling method)};\\
b_2&=-0.024(6) \quad \text{(heating method)};\\
b_2&=-0.025(9) \quad \text{(overlap method)};
\end{aligned}
\end{equation}
while more recent results are:
\begin{equation}\label{lattice results for b2 (2)}
\begin{split}
b_2 & = -0.026(3) \qquad \text{(see Ref. \cite{b2-lattice_1})};\\
b_2 & = -0.0216(15) \quad \text{(see Ref. \cite{b2-lattice_2})}.
\end{split}
\end{equation}
Starting from our results for the topological susceptibility and for the second cumulant in the various cases described, we find:
\begin{itemize}
\item \emph{Chiral Effective Lagrangian $\mathcal{O}(p^2)$}:
\begin{eqnarray}\label{Eff Ch Lagr: numerical b_2}
b_2^{(L=2)} =& -0.026(1) \nonumber \\
b_2^{(L=3)} =& -0.025(1)
\end{eqnarray}
\item \emph{ENL$_{\sigma}$ model}:
\begin{equation}\label{ELsm: numerical b_2}
b_2^{(L=3)}=-0.0222(1)
\end{equation}
\item \emph{WDV model}:
\begin{equation}\label{WDV: numerical b_2}
b_2^{(L=3)}=-0.0222(4)
\end{equation}
\item \emph{Interpolating model}:
\begin{equation}\label{IM: numerical b_2}
b_{2~(\omega_1=1)}^{(L=3)}=-0.0222(4)
\end{equation}
\end{itemize}
In figure 2, these theoretical estimates for $b_2$ (for the \emph{full}
theory with $L=3$) are reported together with the above-mentioned
lattice (pure-gauge) results.
\begin{figure}[!ht]
\begin{center}
\centering
\includegraphics[width=12 cm]{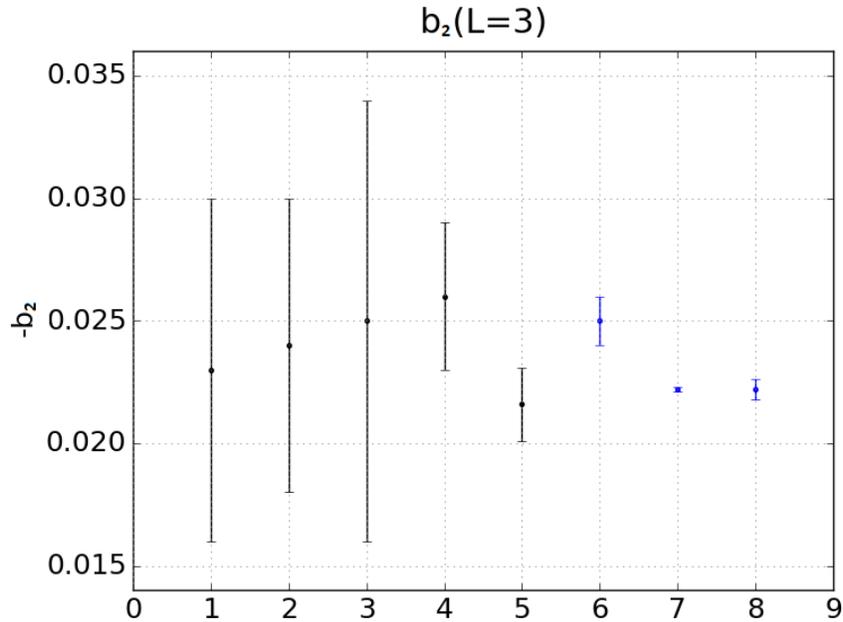}
\end{center}
\caption{The five lattice (pure-gauge) results, reported in Eqs.
\eqref{lattice results for b2 (1)} and \eqref{lattice results for b2 (2)},
and the three theoretical estimates for the \emph{full} theory with $L=3$,
reported in Eqs. \eqref{Eff Ch Lagr: numerical b_2},
\eqref{ELsm: numerical b_2}, and \eqref{WDV: numerical b_2}-\eqref{IM: numerical b_2}, are shown (from left to right).}
\end{figure}
We notice that the lattice (pure-gauge) results turn out to be compatible (in
almost all cases) with our theoretical estimates: this global accordance is
quite impressive, considering that our results have been derived in \emph{full}
QCD rather than in a pure Yang-Mills theory.
We also recall that, on the basis of the results obtained in Secs.
3.1, 4.1, and 5.1, the value of the ratio $b_2$ tends, in the infinite
quark-mass limit, to the pure-gauge value $b_2^{(YM)} = -\frac{1}{12} \simeq -0.083$ (also obtained using a pure-gauge instanton-gas model) in the $ENL_\sigma$
model [see Eq. \eqref{ELsm: pure-gauge limit of chi and c4}], while it tends
to the pure-gauge value $b_2^{(YM)} = 0$ in the $WDV$ model (and in the
\emph{Interpolating} model with $\omega_1 = 1$)
[see Eq. \eqref{WDV: pure-gauge limit of chi and c4}]: therefore, we see
that both the lattice pure-gauge data and our \emph{full}-QCD theoretical
estimates lie in between these two different values and (considering the errors) they disagree with both of them, even if they are considerably closer to the
second one.
It will be interesting to see if future more precise lattice data
(including also the effects of quarks with physical masses) will confirm
(or not) this curious coincidence.

\newpage

\renewcommand{\Large}{\large}

\end{document}